\documentclass[aps,prb,reprint,longbibliography,superscriptaddress]{revtex4-2}
\usepackage{mathrsfs}
\usepackage{amsmath}
\usepackage{amsfonts}
\usepackage{amssymb}
\usepackage{amsthm}
\usepackage{graphicx}
\usepackage{natbib}
\usepackage{color}
\usepackage{hyperref}
\usepackage{bm}
\usepackage[caption=false]{subfig}
\usepackage{verbatim}
\usepackage{ulem}

\usepackage{calrsfs}
\DeclareMathAlphabet{\pazocal}{OMS}{zplm}{m}{n}
\newcommand{\Aa}{\mathcal{A}}
\newcommand{\Ba}{\mathcal{B}}
\newcommand{\Ca}{\mathcal{C}}

\providecommand{\U}[1]{\protect\rule{.1in}{.1in}}

\begin{document}
\title{Hybrid collective excitations in topological superconductor/ferromagnetic insulator heterostructures}

\author{T. Karabassov}
\email{iminovichtair@gmail.com}
	\affiliation{Moscow Institute of Physics and Technology, Dolgoprudny, 141700 Moscow region, Russia}
    \affiliation{HSE University, 101000 Moscow, Russia}

\author{I.V. Bobkova}
\email{ivbobkova@mail.ru}
\affiliation{Moscow Institute of Physics and Technology, Dolgoprudny, 141700 Moscow region, Russia}
\affiliation{HSE University, 101000 Moscow, Russia}

\author{A.M. Bobkov}
\affiliation{Moscow Institute of Physics and Technology, Dolgoprudny, 141700 Moscow region, Russia}

\author{A.S. Vasenko}
\affiliation{HSE University, 101000 Moscow, Russia}

\author{A.A. Golubov}
\affiliation{Moscow Institute of Physics and Technology, Dolgoprudny, 141700 Moscow region, Russia}
\affiliation{HSE University, 101000 Moscow, Russia}

\begin{abstract}

We develop a linear response theory for the dynamical proximity effect in topological superconductor/ferromagnetic insulator (TS/FI) hybrids. Our approach integrates the nonequilibrium quasiclassical Keldysh–Usadel formalism for the TS with the Landau–Lifshitz–Gilbert equation for the FI's magnetization dynamics. This framework reveals a proximity-induced coupling between magnons and superconducting collective modes. Crucially, we find that spin–momentum locking in the TS surface state drives a hybridization between magnons and the superconducting Nambu–Goldstone (phase) mode, giving rise to composite magnon–Nambu-Goldstone excitations. We analyze the coupling strength's dependence on key parameters both analytically and numerically. In contrast, we demonstrate that the Higgs (amplitude) mode does not couple to magnons at linear order and is thus excluded from the hybrid excitation spectrum. The hybridization between magnons and the superconducting phase mode provides a mechanism for the interconversion of spin signals and the spinless signals carried by collective superconducting excitations, thereby giving new impetus to the development of superconducting spintronics.
\end{abstract}

\maketitle

\newpage

\section{Introduction}

The coupling between the superconducting condensate and non-superconducting excitations via the proximity effect is of significant interest, both from fundamental and applied perspectives. In particular, the collective excitation of the magnetic system—a magnon—represents a promising platform for low-dissipation spintronics \cite{pirro2021advances}. Such coupling can occur via both electromagnetic interactions \cite{Yu2025_review} and the proximity effect. In low-dimensional superconducting systems, the dominant mechanism is the proximity effect.

The static proximity effect in superconductor/ferromagnet (S/F) structures is well studied. A plethora of phenomena arising from proximity-induced interactions have been discovered in S/F hybrid systems, including the superconducting spin-valve effect \cite{tagirov1999,baladie2001,fominov2003,mironov2014,oh1997,bovzovic2005,bobkov2026,cadden2008,deutscher1969,nowak2008,gu2002,moraru2006,gu2015,westerholt2005,di2019,ianovskaia2025, Zdravkov2013}, the conversion of spin-singlet to spin-triplet states \cite{Buzdin2005,Bergeret2005}, Josephson $0$–$\pi$ transitions \cite{ryazanov2001} and anomalous $\varphi_0$ ground state \cite{Buzdin2008,Bobkov2022,Bobkova_2022review,Bobkov2024_chains,Jeon2022,Jeon2025}, the superconducting diode effect \cite{Devizorova2021, nadeem2023superconducting, Karabassov2023, karabassov2023phase, Karabassov2022}, and many others; see \cite{Linder2015, Eschrig2015} for reviews.

On the other hand, when the system becomes time-dependent, the dynamical properties of the magnetic and superconducting subsystems can be significantly modified in S/F hybrid structures \cite{Konschelle2009,Shukrinov2017,Nashaat2019,Chudnovsky2016,Rabinovich2019,Guarcello2020,Bobkova2020,Bobkov2024_modes,Bobkov2024_voltage,Johnsen2021,Bobkova2021,Bobkov2021,Bobkova2018,Takashima2017,Halterman2016,Linder2011,Holmqvist2011,Nussinov2005,Zhu2004}. For instance, due to partial singlet–triplet conversion, composite quasiparticles consisting of magnons and clouds of triplet Cooper pairs can emerge at the interface between a superconductor and a ferromagnetic (FI) or antiferromagnetic (AFI) insulator \cite{Bobkova2022,Bobkov2023}. In addition to the partial conversion of singlet Cooper pairs into triplet states via proximity to a magnetic system, the superconducting condensate possesses another important property: the ability to sustain collective modes.

Even in the most basic case of s-wave pairing the superconductor has two collective modes associated with order parameter (OP) excitation: phase (or Nambu-Goldstone) and amplitude (or Higgs) modes\cite{Anderson1958,Anderson1958_2,Nambu1960,Goldstone1961,Arseev2006,Anderson1963,Varma2002,Pekker2015,Nosov2024,Kurkjian2019,Dzero2025, volkov1973collisionless,Derendorf2024,Shimano2020}. It should be mentioned that superconducting systems with more complex gap structure can hold more complicated collective modes, including Leggett modes in multi-band superconductors \cite{leggett1966number, Sun2020, hu2024,giorgianni2019leggett, Bittner2015,Anishchanka2007}, clapping modes in unconventional superconductors\cite{Wolfle1976, Tewordt1999,poniatowski2022,Matsushita2022}, Bardasis-Schrieffer mode in systems with subdominant pairing potential\cite{Bardasis1961,Maiti2015,Maiti2016}. Furthermore, theoretical predictions suggest that in the topological superconductors with a nematic order parameter, chiral Higgs and nematicity mode may also emerge\cite{Uematsu2019}. 
Recently it was reported that in  S/F hybrid structures in the presence of spin-orbit coupling (SOC) the Higgs mode can be induced by the magnetization dynamics via the linear  coupling  process \cite{Lu2022,Plastovets2023,Silaev2020_2}. 


Typically, the phase mode is excluded from analysis in three-dimensional (3D) superconductors. This is because, in the presence of long-range Coulomb interactions, the phase mode hybridizes with the gapped 3D plasmon. This hybridization lifts the phase mode's energy up to the plasma frequency, making it indistinguishable from plasma oscillations \cite{Anderson1963}. An important exception occurs near the superconducting critical temperature, where a large population of thermally-excited quasiparticles can screen the charge. This allows a gapless phase mode, known as the Carlson-Goldman (CG) mode, to emerge \cite{Carlson1975,Schmid1975,Artemenko1975,Artemenko1979,Sun2020,langenberg1986nonequilibrium}. The interplay between plasmons and the CG mode in thin films has been a subject of theoretical study \cite{Mooij1985}. 

Crucially, the situation is fundamentally different in two-dimensional (2D) systems. In 2D, the plasmon itself becomes a gapless mode, with a dispersion that vanishes as momentum goes to zero \cite{Vignale2005}. Therefore, when this gapless 2D plasmon hybridizes with the superconducting phase mode, it cannot impart a gap. Consequently, the phase mode in a 2D superconductor remains gapless at all temperatures \cite{ohashi1998, Buisson1994,Kliewer1967,Sun2020,Karuzin2025}.
Thus, considering the rapid advancements of superconducting 2D materials and topological surface physics \cite{sato2017,zollner2024}, exploring the phase mode in 2D systems becomes essential for further understanding of fundamental physical mechanisms as well as possible applications in the field of low dissipation electronics. Despite the achieved progress in the studies of the dynamical proximity effects, the coupling between magnons and superconducting phase Nambu-Goldstone (NG) modes have not yet been reported in the literature. 

In this work we focus on the examination of the mutual influence of the superconducting collective modes and magnon excitations in topological superconductor/ferromagnetic insulator (TS/FI) hybrid structures. The conductive surface state of the TS represents a basic system sustaining the strongest SOC in the form of the full spin-momentum locking \cite{Burkov2010,Culcer2010,Yazyev2010,Li2014}. 
The interaction of magnons with surface plasmons of magnetic insulators and nonsuperconducting topological insulators was already discussed \cite{Efimkin2021,Dyrdal2023}. Furthermore, the interaction of the superconducting plasmon in a one-dimensional resonator and magnons in an antiferromagnetic insulator was also discussed theoretically \cite{Malshukov2021}. Here, we develop a linear response theory of collective excitations in 2D superconducting systems having the property of the full spin-momentum locking contacted with a thin-film ferromagnetic insulator. The theory is developed in the framework of the non-equilibrium Keldysh quasiclassical approach. 
We show that the spin-momentum locking leads to hybridization of the NG phase mode with the magnon and the appearance of composite excitations consisting of a magnon in the FI accompanied by oscillations of the phase of the superconducting order parameter in TS. The dependence of the coupling strength on the relevant physical parameters is studied analytically and numerically. At the same time, it is demonstrated that the Higgs mode is not coupled to magnons in the linear order and does not  form hybrid collective excitations with it.

The paper is organized as follows. Section~\ref{sec:model} formulates the studied model. Section~\ref{sec:approach} presents a qualitative picture of the effect and details the construction of the theoretical framework. The resulting spectra of both uncoupled and coupled superconducting modes and magnons are presented in Sec.~\ref{sec:spectrum}, and our conclusions are summarized in Sec.~\ref{sec:conclusions}. Further supporting details are provided in the appendices: Appendix~\ref{sec:appendix_A} covers technical aspects of the Green's function calculations, and Appendix~\ref{app:amplitude_response} presents symmetry relations used to prove the absence of magnon-Higgs mode interaction.


\section{System and model}
\label{sec:model}
\begin{figure*}[t]
\includegraphics[width=2\columnwidth]{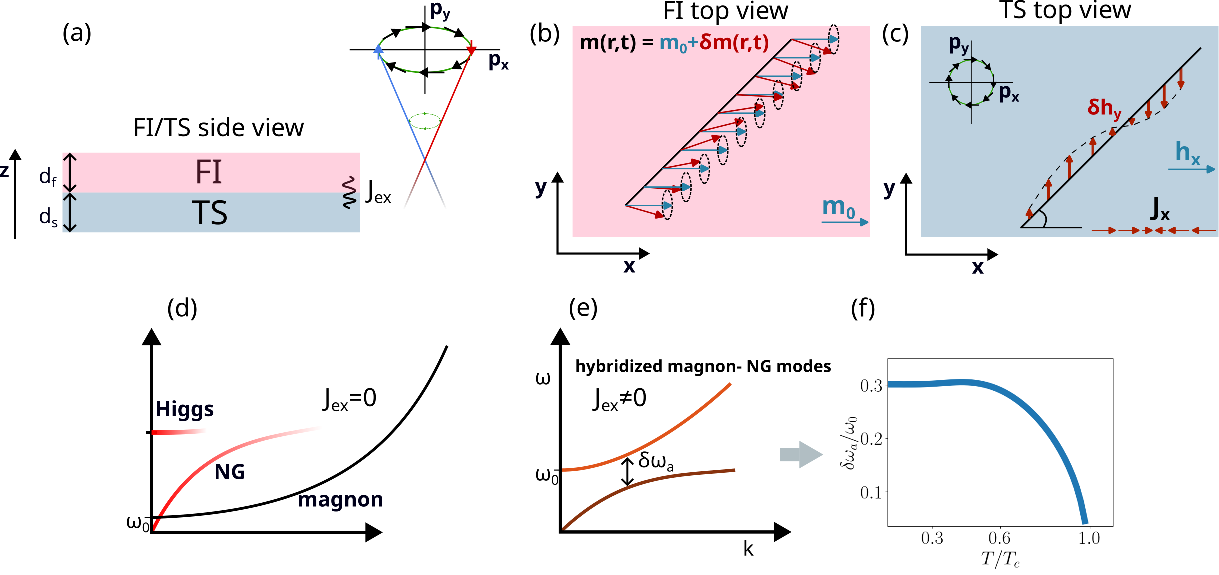}
\caption{Schematic of the TS/FI heterostructure. (a) Side view of the system, illustrating the interface exchange coupling $J_{ex}$ between the superconducting and magnetic subsystems. Normal-state dispersion relation of the 2D conductive surface state of the TI, showing spin-momentum locking is depicted at the top right corner. The equilibrium magnetization of the FI $\bm m_0 = \hat x$ and a magnon excitation $\delta \bm m$ of the FI induce in the  TS an equilibrium effective exchange field $\bm h_x$ and a magnon-induced dynamic contribution $\delta \bm h_y$, respectively. The dynamic exchange field $\delta \bm h_y$  triggers a time-dependent response of the superconducting order parameter in the TS, resulting in a hybridized magnon–Nambu–Goldstone (NG) excitation. (b) Top view of the FI layer. (c) Top view of the TS layer.  (d) The basic modes of the non-interacting system (i.e. $J_{ex} =0$).
(e) Schematic plot of the magnon -NG mode hybridization in the system and (f) anticrossing strength as a function of the temperature.}
 \label{fig:sketch}
\end{figure*}
We consider a topological superconductor/ferromagnetic insulator (TS/FI) heterostructure, see Fig.~\ref{fig:sketch}(a)-(c). On the surface of the TS, a 2D superconducting state with full spin-momentum locking occurs. The corresponding normal state electron dispersion is shown in Fig.~\ref{fig:sketch}(a). The normal state Fermi surface of the conductive surface state of the TS is represented by  the only helical band \cite{Zhang2018,Hao2019} similar to the conductive surface state of topological insulator \cite{Burkov2010,Culcer2010,Yazyev2010,Li2014}.  The electron spin at the Fermi surface always makes the right angle with its momentum with a definite helicity. 

The FI layer induces an exchange field in the TS underneath via the proximity effect \cite{Bergeret2018,Hao1991,Moodera2007,Cottet2009,Eschrig2015bc}. In the framework of the interface exchange model \cite{Bobkova2021}, which works well for interfaces with ferromagnetic insulators, the induced exchange field in the TS takes the form $\bm h = -J_{ex} M_s \bm m / (2 \gamma d_s)$, where $J_{ex}$ is the interface exchange coupling constant, $M_s$ is the saturation magnetization of the FI, $\gamma$ is the gyromagnetic ratio magnitude, $\bm m$ is the unit vector along the FI magnetization and $d_s$ is the effective thickness of the TS surface conductive layer, which in the considered case is of the order of a few interatomic lengths. 

The resulting effective Hamiltonian of electrons in the TS 2D surface conductive  layer takes the form:
\begin{align}
H&=\int d^2 r \Biggl\{\Psi^\dagger (\bm r,t)\bigl[-i \hbar v_f(\bm \nabla_{\bm r}\times \hat z)\bm \sigma  \nonumber \\
&- \mu + e \phi(\bm r)+ V_{imp}(\bm r)- 
\bm h \bm \sigma \bigr]\Psi(\bm r,t) \nonumber \\
&+\Delta(\bm r,t)\Psi^\dagger_\uparrow (\bm r,t) \Psi^\dagger_\downarrow (\bm r,t) + \Delta^*(\bm r,t)\Psi_\downarrow (\bm r,t) \Psi_\uparrow (\bm r,t) \Biggr\},~~~~
\label{ham}
\end{align}
where $\Psi^\dagger(\bm r,t)=(\Psi^\dagger_\uparrow(\bm r,t),\Psi^\dagger_\downarrow(\bm r,t))$ is the electron creation operator, $\hat z$ is the unit vector normal to the surface of TS, $v_f$ is the electron Fermi velocity, $\mu$ is the chemical potential, $\phi(\bm r)$ is the scalar electric potential and $\Delta(\bm r,t)$ is the superconducting order parameter (OP), which is assumed to be of $s$-wave singlet type and should be calculated self-consistently as $\Delta(\bm r,t) = \lambda \langle \Psi_\downarrow \Psi_\uparrow \rangle$.  $\bm \sigma = (\sigma_x, \sigma_y, \sigma_z)$ is a vector of Pauli matrices in spin space. The term $V_{imp}(\bm r)$ includes the nonmagnetic impurity scattering potential $V_{imp}(\bm r)=\sum \limits_{\bm r_i}V_i \delta(\bm r - \bm r_i)$, which is of a Gaussian form $\langle V_{imp}(\bm r)V_{imp}(\bm r')\rangle = (1/\pi \nu \tau)\delta(\bm r - \bm r')$ with $\nu=\mu/(2\pi v_f^2)$. 

Our goal is to consider {\it the dynamical coupling} between the magnons and the superconductive collective modes. For this reason the magnetization in the FI is assumed to be time-dependent and takes the form $\bm{m}(\bm r, t) = \bm m_0 + \bm{\delta m} (\bm r,t)$, where $\bm m_0 = \bm \hat x$ is the equilibrium magnetization and $ \bm {\delta m}(\bm r,t) = {\rm Re}[\delta \bm m e^{i(\bm{k \cdot r} + \omega_r t)}] \exp{(- \kappa t)}$ accounts for the spin wave, where $\omega_r$ is the magnon frequency and $\kappa$ characterizes the decay of the spin wave, which originates from the intrinsic damping of the magnet and also from the coupling to the superconducting excitations. $\bm{m}(\bm r, t)$ obeys the Landau-Lifshits-Gilbert equation (LLG), see Sec.~\ref{subsec:magnon}. 

The induced in the TS exchange field takes the form $\bm h = \bm h_0 + \delta \bm h$, where $\bm h_0 = -J_{ex} M_s \bm m_0 / (2 \gamma d_s) = h_0 \hat x$ is the equilibrium exchange field. We consider the limit $\mu \gg  (h,\Delta)$, when the Fermi level of the conductive surface state of TS is far from the Dirac point. In this limit the out-of-plane $z$-component of the magnon excitation produces negligible $z$-component of the  effective exchange field, which is $\sim (h_0/\mu) m_z $ \cite{Zyuzin2016,Bobkova2016}, and we disregard it. Then the magnon-induced effective exchange field $\delta \bm h$ takes the form 
\begin{align}
&\delta \bm{h} =  \delta h \cos\left(\omega_r t + \bm k \cdot \bm r\right) e^{-\kappa t} \hat y = \nonumber \\
&(\delta \bm h_{\omega_r, \bm k} e^{i \left(\omega_r t + \bm k \cdot \bm r \right)} + \delta \bm h_{-\omega_r, -\bm k} e^{-i \left(\omega_r t + \bm k \cdot \bm r \right)})e^{-\kappa t},
\label{delta_h}
\end{align}
where $\delta \bm h_{\omega_r, \bm k} = \delta \bm h_{-\omega_r, -\bm k} = (\delta h/2) \hat y$.
If one assumes a spatially homogeneous order parameter  in the equilibrium state of the system $\Delta (\bm r) = \Delta$, then in this state a spontaneous electric current occurs \cite{Yip2002}. It was shown that such an equilibrium state does not have a minimal energy. The ground state of the system in the presence of the equilibrium exchange field $\bm h_x$ is the so-called helical state \cite{Edelstein1989,Barzykin2002,Samokhin2004,Kaur2005,Dimitrova2007,Houzet2015,Mineev2011} and is described by the phase-inhomogeneous order parameter $\Delta(\bm r) = \Delta \exp[i \bm q \bm r]$, where $\Delta$ is the absolute value of the OP at a given temperature $T$, and $\bm q = -2h_0 \hat e_y/ \hbar v_f$ is determined from the condition that the total current in the ground state is zero. The superconducting collective modes are described by the time dependent correction to the OP, that is in general $\Delta (\bm r, t) = \Delta \exp[i \bm q \bm r] + \delta \Delta (\bm r, t)$.

\section{Qualitative picture and theoretical formalism}
\label{sec:approach}

This section outlines the qualitative behavior of collective excitations and presents the theoretical framework for calculating their spectrum in the TS/FI hybrid system, based on the non-equilibrium quasiclassical Keldysh formalism.

The spectrum is determined by coupled fluctuations of the magnetization in the FI, the complex-valued order parameter in the TS, and the scalar potential, the latter being necessary to account for charge oscillations. The observables corresponding to these fluctuating fields are derived from the Keldysh Green's functions in the linear response approximation, as detailed in Sec.~\ref{subsec: GF_derivation}. The dynamics are governed by three key equations: the self-consistency equation for superconducting gap oscillations ( Sec.~\ref{subsec: GF_derivation}), the Poisson equation for the self-consistent scalar potential under Coulomb interaction (Sec.~\ref{subsec:scalar}), and the LLG equation for magnon dynamics in the FI (Sec.~\ref{subsec:magnon}). These equations must be solved simultaneously to obtain the full spectrum as well as damping (decay rate) of the collective modes (Sec.\ref{subsec:total_spect}), as summarized in Table~\ref{table}.

The spin-momentum locking provides a linear response of the OP to $\delta \bm h$: 
\begin{align}
\delta \Delta(\bm r, t) = (\delta {\Delta}_{\omega_r,k} e^{i \left(\omega_r t + \bm k \cdot \bm r \right)} + \nonumber \\
\delta {\Delta}_{-\omega_r,-k} e^{- i \left(\omega_r t + \bm k \cdot \bm r \right)})e^{-\kappa t}.~~~~
\end{align}

It is convenient to decompose the dynamical correction to the OP into contributions from the amplitude (Higgs) mode and the phase (NG) mode, defined respectively as
$\delta \Delta_{a} =[\delta \Delta_{\omega_r, \bm k} + \delta \Delta_{-\omega_r, -\bm k}^* ]/2$ and $\delta \Delta_{p } =[\delta \Delta_{\omega_r, \bm k} - \delta \Delta_{-\omega_r, -\bm k}^* ]/2i$. The corresponding linear response corrections can then be written in the form:
\begin{eqnarray}
\delta \Delta^{a} = F_{\omega_r,k}^a(\hat {\bm k} \times \delta \bm h_{\omega_r, \bm k})\hat z , \nonumber \\
\delta \Delta^{p} = F_{\omega_r,k}^p(\hat {\bm k} \times \delta \bm h_{\omega_r, \bm k})\hat z .
\label{op_magnon}
\end{eqnarray}
Here $\hat {\bm k} = \bm k/|k|$ and $F_{\omega_r,k}^{a,p}$ are calculated microscopically using the Keldysh Green’s function formalism within the framework of the Usadel equations (See Sec. \ref{subsec:scalar} and Sec.\ref{subsec:total_spect}). 

It is important to note that in the absence of spin–momentum locking, the superconducting order parameter does not couple linearly to magnons\cite{Bobkova2022}. In S/FI hybrids based on conventional superconductors without SOC, only a linear-response {\it triplet} correction to the superconducting correlations arises due to magnon excitations.
As a result, magnons  can be dressed by triplet pairs forming magnon-cooparons \cite{Bobkova2022}, but composite excitations of magnons and superconducting collective modes do not occur because of the absence of a linear response {\it singlet} correction.

In the present case, the linear coupling arises directly from the fact that, in a TS, the superconducting correlations consist of mixed-spin components with equal singlet and triplet amplitudes \cite{Bobkova2017}. A second key observation is that $F_{\omega_r,k}^{a} = 0$, as shown by microscopic calculations presented in  App. \ref{app:amplitude_response}. This result implies that magnons do not excite the amplitude (Higgs) mode in a topological superconductor characterized by full spin–momentum locking and residing in its helical ground state.

On the other hand, the superconducting phase NG mode generates  {\it ac} current response in the superconductor. The magnon-induced {\it ac} effective exchange field $\delta \bm h$ also produces {\it ac} electric current in the TS via the inverse magnetoelectric effect \cite{Shen2014,Ganichev2002,Sanchez2013,Sanchez2016,Zhang2016,Yip2002,Bobkova2004,Mironov2017,Pershoguba2015,Malshukov2016,Malshukov2020,Malshukov2020_2}. The total {\it ac} current takes the form
\begin{eqnarray}
\bm J = (\bm J_{\omega_r,k} e^{i \left(\omega_rt + \bm k \cdot \bm r \right)} +  \bm J_{\omega_r, \bm k}^* e^{-i \left(\omega_r t + \bm k \cdot \bm r \right)})e^{-\kappa t}. 
\label{current_magnon}
\end{eqnarray}
Up to the linear order with respect to the OP perturbations and $\delta \bm h$ it can be expanded as: 
\begin{eqnarray}
\bm J_{\omega_r, \bm k}  = \bm J_{p} \delta \Delta^{p} + \bm J_{\delta h} \delta \bm h_{\omega, \bm k} ,
\label{current_expansion}
\end{eqnarray}
where $\bm J_{p} \propto \hat{\bm k}$ physically represents the current response of the TS on the NG mode. The current response to the amplitude mode is absent. $\bm J_{\delta h}$ contains two terms: $\bm J_{1,\delta h} \propto  (\delta \bm h_{\omega, \bm k} \times \hat z)$ and $\bm J_{2,\delta h} \propto \hat {\bm k}([\hat {\bm k} \times \delta \bm h]\hat z)$. The current response functions $\bm J_{p}$ and $\bm J_{\delta h}$ are calculated microscopically (See Sec.\ref{subsec:magnon}). A key property of the TS is that an electric current flowing through it induces an electron spin polarization $\bm s$ via the direct magnetoelectric effect. The general expression for this effect, which holds in both the superconducting and normal states, is given by\cite{Shiomi2014,Bobkova2016}:
\begin{equation}
     \bm s = -\frac{1}{2 e v_f} (\hat z \times \bm J).
     \label{polarization_current}
\end{equation}
 The dynamics of spin wave in FI is described by the LLG equation:
\begin{eqnarray}
\frac{ \partial \bm{m}}{\partial t} = - \bm{m} \times \left(D_m \bm{\nabla}^2 \bm{m} + \gamma \bm {H_{eff}}\right)  + \nonumber \\
\alpha \bm{m} \times \frac{\partial \bm{m}}{\partial t} + \frac{J_{ex}}{d_f} \bm{m} \times \bm{s}.
\label{LLG}
\end{eqnarray}
In this  equation $\bm{H_{eff}} = K m_x \hat x$, where  $K$ is the uniaxial anisotropy constant, $D_m$ is the magnon stiffness and $\alpha$ is the  Gilbert damping parameter. The last term represents the spin torque, which accounts for the back-action of the TS on the FI via the interfacial exchange interaction between the FI magnetization and the spin polarization $\bm s$ in TS. The factor $1/d_f$, where $d_f$ is the thickness of the FI layer, arises from averaging the interface exchange term over the FI thickness.
Thus, in the TS, the NG superconducting mode generates an {\it ac} spin polarization, which in turn can excite magnons in the FI via the torque term in the LLG equation.

The nonzero response of the OP on the magnon, described by Eq.~(\ref{op_magnon}) together with the nonzero response of the FI magnetization on the NG mode described by the LLG equation (\ref{LLG}) and Eq. \eqref{polarization_current} results in the appearance of composite excitations of magnons and the NG mode in FI/TS heterostructures. The hybrid excitations can be investigated in the basis $\hat \Psi = (\delta \Delta_p, \delta h)^T$, where $\delta h$ is directly proportional to $\delta m_y$ via $\delta m_y = -2\gamma d_s \delta h/(J_{ex})M_s$. The  spectrum of the hybridized excitations can be found from the combination of the linearized self-consistency equation for the OP and the linearized LLG equation, which can be written as follows (See Sec. \ref{subsec:magnonNG} for the derivation and definition of $M^p$):
\begin{equation}
    \begin{pmatrix}
        M^p & M^p F^p \\
        c_s (\bm J_p \cdot \hat x) &  M^h
    \end{pmatrix} \hat \Psi_p =0 ,
    \label{matrix_modes_qu}
\end{equation}
The upper line of Eq.~(\ref{matrix_modes_qu}) is the linearized self-consistency equation for the superconducting gap and the bottom  line comes from the LLG equation with  $c_{s} = J_{ex}h_0 /( e v_f d_f)$ and $M^h = -(\omega_b + i \alpha [\omega_r + i \kappa ]) + c_s  (\bm J_{\delta h} \cdot \hat x) + [\omega_r + i \kappa ]^2/(\omega_b + i \alpha \omega)$.

In the absence of the interface exchange coupling $J_{ex}=0$, that is $ \delta h = 0$, equation $ M^h = 0$ determines the bare magnon modes $\omega_b = \gamma K + D_m k^2$ and $ M^p = 0$ provides the phase NG mode of the charged 2D superconductor \cite{Sun2020,Buisson1994,Karuzin2025} with dispersion relation
\begin{align}
\hbar \omega_{NG}= \pi \sqrt{2 \Delta_0  k_B T_c \left( \frac{\hbar^2 \omega_p^2k\xi}{4 \pi^2 \Delta_{0} k_B T_c}   + ( k \xi)^2\right)},
\label{NG_analytical}
\end{align}
where $\omega_p = \sqrt{4 \pi^2 e^2 \nu \Delta_0 D/ \hbar\xi} = \sqrt{2 \pi^2 \sigma d_s \Delta_0/(\hbar\xi)}$ is the superconducting plasma frequency with $D$ being the diffusion constant of the TS. The spectrum of the NG mode is schematically shown in Fig. \ref{fig:sketch} (d). 
\begin{table}[t]\label{table_1}
   \caption{Problem description} 
   \label{table}
   \small
   \begin{tabular}{lccr}
   \toprule\toprule
   \textbf{MF potential} & \textbf{Equation} & \textbf{Observable}\\ 
   $\delta \Delta$ & Gap equation & Superconducting gap $\Delta$ \\
   $\delta \phi$ & Poisson equation & Electron density $\rho$\\
   $\delta \bm m$ & LLG equation & Spin density $\bm s$
   \end{tabular}
\end{table}

At $J_{ex} \neq 0$ the NG mode and the magnon are coupled due to nonzero $c_s$ and $F^p$ in Eq.~(\ref{matrix_modes_qu}). The coupled modes are represented in Fig.~\ref{fig:sketch} (e). Due the spin-momentum locking which dictates the symmetry of the linear response in the form $(\hat {\bm k} \times \delta \bm h_{\omega, \bm k})\cdot \hat z$ the coupling is anisotropic. Its magnitude is maximal when the magnon propagates along the equilibrium magnetization $\hat {\bm k} = \hat x$, leading to the anticrossing between the NG and the magnon modes, and is zero if $\hat {\bm k} = \hat y$.  Thus, the eigenmodes of the TS/FI bilayer become composite excitations magnon - Nambu-Goldstone. The analytical expressions for the anticrossing strength are derived in Sec. \ref{subsec:magnonNG}. However, even without doing microscopic calculations from Eq.~(\ref{matrix_modes_qu}) it is seen that the strength of interaction between the magnon and the NG mode, which can be quantified by $\delta \omega_a  = \omega_{up}(k_i) - \omega_{dn}(k_i)$ (See Fig. \ref{fig:sketch} e), scales linearly with the interface exchange coupling $\delta \omega_a \propto \sqrt {M^p F^p c_s (\bm J_p \cdot \hat x)} \propto h_0$ since $c_s \propto h_0^2$ and the other involved quantities do not contain $h_0$. The dependence of $\delta \omega_a$ on temperature is illustrated schematically in Fig.~\ref{fig:sketch}(f). It tends to zero at $T \to T_c$ indicating that the effect is of purely superconducting nature and strongly differs from the nonsuperconducting magnon-plasmon coupling of the magneto-electric origin, which was reported in \cite{Efimkin2021, Dyrdal2023}.

\subsection{Excitation-induced first-order corrections to the electronic Green's function}
\label{subsec: GF_derivation}
In the considered case $(h, \Delta) \ll \mu$ the 2D conductive surface state of the TS can be described in the framework of the quasiclassical approximation of the Green's functions approach. In this work we assume the diffusive limit, i. e.  the elastic scattering length $l \ll \xi$, where $\xi = \sqrt{\hbar D/2 \pi T_c}$ is the superconducting coherence length and $D$ is the diffusion constant. Then the system can be described in terms of the Usadel equation \cite{Zyuzin2016,Bobkova2016} for the quasiclassical Green's function $\check {\check g}(\bm n_F, \bm r, \varepsilon,t)$, which is a $8 \times 8$ matrix in a direct product of the particle-hole, spin and Keldysh spaces. The quasiclassical Green's function depends on the quasiparticle energy $\varepsilon$, 2D radius-vector $\bm r$ in the TS surface plane and on the direction of the electron trajectory determined by the unit vector $\bm n_F=\bm p_F/p_F=(n_{F,x},n_{F,y},0)$, where $\bm p_F$ is the  electron momentum at the Fermi surface. Since we consider a non-stationary problem dealing with excitations, the Green's function also depends on time $t$. The  spin structure of the quasiclassical Green's function  is dictated by the projection onto the conduction band of the TI surface states:
\begin{equation}
\check {\check g}(\bm n_F, \bm r, \varepsilon,t)= \check g(\bm r, \varepsilon,t)\frac{(1+\bm n_\perp \bm \sigma)}{2},
\label{spin_structure}
\end{equation}
where  and $\bm n_\perp=(n_{F,y},-n_{F,x},0)$ is a unit vector perpendicular to the quasiparticle trajectory and directed along the quasiparticle spin, which is locked to the quasiparticle momentum. $\check g$ is a {\it spinless} $4 \times 4$ matrix Green's function in the particle-hole and Keldysh spaces, which describes mixed singlet-triplet correlations in the system and in the diffusive limit is isotropic in the momentum space.  Its explicit structure in the Keldysh space takes the form: 
\begin{equation}
\Check{g} =
\begin{pmatrix}
\hat{g}^R & \hat{g}^K \\
0 & \hat{g}^A
\end{pmatrix},
\label{keldysh}
\end{equation}
where $\hat g^{R,A,K}$ are retarded, advanced and Keldysh components of the Green's function, and each of them is a $2 \times 2$ matrix in the particle-hole space. 
The Usadel equation for spinless Green's function in the mixed representation $(\varepsilon,t)$ takes the form:
\begin{align}
&i \hbar D \hat{\nabla} \left(\Check{g} \otimes \hat{\nabla} \Check{g}\right) = \left[\check{\Lambda} + i \check{\Delta}(\bm r)  - \delta \check \phi (\bm r),
\check{g}\right]_\otimes, \nonumber \\
&\check{\Lambda} = \begin{pmatrix}
    \hat{\Lambda}^R & \hat{\Lambda}^K \\
    0 & \hat{\Lambda}^A
\end{pmatrix} = \begin{pmatrix}
    (\epsilon + i \Gamma) \tau_z & 2 i \Gamma \tau_z \tanh{\left(\frac{\epsilon}{2 T}\right)} \\
    0 & (\epsilon - i \Gamma) \tau_z
\end{pmatrix} , \nonumber \\
& i \hat{\Delta}(\bm r) = i
\begin{pmatrix}
0 & \Delta(\bm r) \\
\Delta^*(\bm r) & 0
\end{pmatrix} \mathrm I_K , \quad \delta \check \phi (\bm r) = e \delta  \phi (\bm r) \tau_0 I_K,
\label{eq:Usadel_general}
\end{align}
where $\mathrm I_K$ is the unit matrix in the Keldysh space, $\hat{\nabla} X = \nabla X + \frac{i}{\hbar v_f} \left[\left(h_x \hat y - h_y \hat x\right)\tau_z, X\right]_\otimes$. $h_x$ and $h_y$ are the in-plane components of a proximity induced exchange field, and $\Gamma$ is the phenomenological Dynes parameter, which takes into account inelastic scattering processes. The circle product $\otimes$ is defined as
\begin{align}
    A(\varepsilon,t)\otimes B(\varepsilon,t) = \nonumber \\
    \exp \left[- \frac{i}{2}\left(\partial^B_\varepsilon \partial^A_t - \partial^A_\varepsilon \partial^B_t \right)\right] A(\varepsilon,t) B(\varepsilon,t).
\end{align}
The Usadel equation (\ref{eq:Usadel_general}) should be supplied by the normalization condition 
\begin{align}
    \check g \otimes \check g = 1 .
    \label{eq:normalization}
\end{align}
At first let us consider the equilibrium state of the system without a magnon. As it was explained above, due to the spin-momentum locking and the proximity-induced exchange field the superconductor is in the helical state. For this reason we consider a phase-inhomogeneous state $\Delta(\bm r) = \Delta_u e^{i \bm q \bm r}$. Then we introduce the unitary transformation $\hat{U} = \exp \left[ i  \bm{q} \cdot \bm{r} \tau_z/2 \right]$. The Usadel equation for the transformed Green's function $\check g_u = \hat U^\dagger \check g \hat U$ takes the form:
\begin{align}
i \hbar D \hat{\nabla}_u \left(\Check{g}_u \otimes \hat{\nabla} \Check{g}_u\right) = \left[ \check{\Lambda} + i \hat{\Delta}_u(\bm r) - \delta \check \phi (\bm r), \Check{g}_u\right]_\otimes, \label{usadel_u}\\
 i \hat{\Delta}_u(\bm r) = i
\begin{pmatrix}
0 & \Delta_u(\bm r) \\
\Delta^*_u(\bm r) & 0
\end{pmatrix} \mathrm I_K
\end{align}
where $\hat{\nabla}_u = \nabla X + i \frac{\bm q}{2}\left[ \tau_z, X \right] + \frac{i}{\hbar v_f} \left[\left(h_x \hat y - h_y \hat x\right)\tau_z, X\right]_\otimes$. From this expression we see that the generalized Cooper pair momentum is $\bm q_{pair} = \bm q + 2(h_x \hat y - h_y \hat x)/\hbar v_f$. The ground state of the system corresponds to zero supercurrent and, consequently, to $\bm q_{pair} = 0$. Therefore, it is a phase-inhomogeneous helical state with $\bm q_{gs} = -2h_x \hat y/v_f = -2h_0 \hat y/\hbar v_f$. In this case the superconducting gap $\Delta$ generates the phase gradient that compensates the spontaneous current driven by $h_0$. 

Now we consider the solution of the Usadel equation in the presence of an excitation. We expand the solution of the Usadel equation and the superconducting order parameter as 
\begin{align}
\check{g}_u &= \check{g}_0 + \delta \check{g}, \label{g_expansion}\\
\hat{\Delta}_u &= \hat{\Delta} + \delta \hat{\Delta},
\label{delta_expansion}
\end{align}
where $\hat{g}_0$ and $\hat{\Delta}$ are the equilibrium expressions and $\delta \hat{g}$ and $\delta\hat{\Delta}$ are the corrections due to the excitation. In the particle-hole space the correction to the Green's function can be written as:
\begin{align}
    \delta \check{g}  = \begin{pmatrix}
    \delta \hat g_{11} & \delta \hat f_{12} \\
    \delta \hat f_{21} & \delta \hat g_{22}
\end{pmatrix},
\end{align}
where $\delta \hat g_{ij}$ and $\delta \hat f_{ij}$ are matrices in the Keldysh space.
Substituting Eqs.~(\ref{g_expansion})-(\ref{delta_expansion}) into the Usadel equation (\ref{usadel_u}), we derive equations for $\delta \hat g$ up to the first order in $\delta h $ and $\delta \Delta$. We put $\bm q = \bm q_{gs}$ because we assume that the ground state of the system corresponds to the helical state. Then for retarded and advanced Green's functions matrices we obtain
\begin{align}
&i \hbar D \left( \hat{g}_0^{R,A} \otimes \nabla^2 \delta \hat{g}^{R,A} - \frac{i}{\hbar v_f} \hat{g}_0^{R,A} \otimes \left[ \partial_x h_y \tau_z, \hat{g}_0^{R,A} \right]_\otimes \right)
\nonumber \\ 
&=\left[ \left (\epsilon \pm i \Gamma \right)\tau_z + i \hat{\Delta}, \delta \hat{g}^{R,A}\right]_\otimes \nonumber \\
&+ i \left[\delta \hat{\Delta}, \hat{g}_0^{R,A} \right]_\otimes - \left[\delta \check \phi, \hat{g}^{R,A}_0\right]_\otimes ,
\label{eq:RA_perturbation}
\end{align}
and for the Keldysh Green's function
\begin{align}
&i \hbar  D\left(\hat{g}_0^R \otimes \nabla^2 \delta \hat{g}^K - \frac{i}{\hbar v_f} \hat{g}_0^R \otimes \left[ \partial_x h_y \tau_z, \hat{g}_0^K \right]_\otimes \right. 
\nonumber \\
&\left. + \hat{g}_0^K \otimes \nabla^2 \delta \hat{g} ^A - \frac{i}{\hbar v_f} \hat{g}_0^K \otimes \left[ \partial_x h_y \tau_z, \hat{g}_0^A \right]_\otimes \right) = \nonumber \\
&\left[\hat{\Lambda}^R + i \hat{\Delta}\right] \otimes \delta\hat{g}^K - \delta\hat{g}^K \otimes \left[\hat{\Lambda}^A+ i \hat{\Delta} \right] 
\nonumber \\
&+ i\left[ \delta \hat{\Delta}, \hat{g}_0^K\right]_\otimes   - \left[ \delta \hat \phi, \hat{g}^K_0\right]_\otimes + C_\Gamma(\bm r,t),
\nonumber \\ 
 &C_\Gamma(\bm r,t) =\left[\hat{\Lambda}^K \otimes \delta\hat{g}^A - \delta\hat{g}^R \otimes \hat{\Lambda}^K \right].
\label{eq:Keldysh_perturbation}
\end{align}
The equilibrium Green's functions take the bulk form:
\begin{align}
    &\hat{g}^{R(A)}_0 = g^{R(A)}_0 \tau_z  + i f^{R(A)}_0 \tau_x, \nonumber \\
    &g^R_0 (\epsilon)= \frac{ sgn(\epsilon) \left(\epsilon + i \Gamma\right)}{\sqrt{(\epsilon + i\Gamma)^2 - \Delta^2}}, \quad f^R_0 (\epsilon)= \frac{\Delta sgn{(\epsilon)}}{\sqrt{(\epsilon + i\Gamma)^2 - \Delta^2}}, \nonumber \\
    &f[g]_0^A = - \left(f[g]_0^R\right)^*, \nonumber \\
    &\hat{g}^K_0 = \left(\hat{g}^R_0 - \hat{g}^A_0\right)\tanh{\left(\epsilon/2T\right)}. 
    \label{eq:bulk}
\end{align}
It is worth noting that in our case, when the Fermi surface is represented by the only helical band and we have the full spin-momentum locking, the effective exchange field fully drops out of the equilibrium Green's function if $\bm q=\bm q_{gs}$.

The response of the superconductor to the magnon described by Eq.~(\ref{delta_h}) should be found in the form:
\begin{equation}
\delta \check{g} = \delta \check{g}_{\omega_r,k} e^{i \left(\omega t + \bm k \cdot \bm r \right)} +\delta \check{g}_{-\omega_r,-k} e^{- i \left(\omega^* t + \bm k \cdot \bm r \right)}.
\label{delta_correction}
\end{equation}
where $\omega \equiv \omega_r +i\kappa$ is the complex value of frequency, including the excitation decay rate $\kappa$. The magnon correction $\delta\Delta$ to the superconducting gap takes an analogous form:
\begin{equation}
\delta \hat \Delta = \delta \hat{\Delta}_{\pm} e^{i \left(\omega t + \bm k \cdot \bm r \right)} +\delta \hat{\Delta}_{\mp} e^{- i \left(\omega^* t + \bm k \cdot \bm r \right)}.
\label{op_correction}
\end{equation}
Here $\delta \hat{\Delta}_\pm = antidiag(\delta \Delta_{\omega_r,k}, \delta \Delta^*_{-\omega_r,-k} )$ and $\delta \hat{\Delta}_\mp = antidiag(\delta \Delta_{-\omega_r,-k}, \delta \Delta^*_{\omega_r,k} )$. Substituting Eqs.~(\ref{delta_correction})-(\ref{op_correction}) into Eqs.~(\ref{eq:RA_perturbation})-(\ref{eq:Keldysh_perturbation}) and taking into account that $h_y = (\delta h/2)e^{i \omega t + \bm k \cdot \bm r}$, for retarded and advanced components we obtain 
\begin{widetext}
\begin{align}
&-  i \hbar  D k^2 \hat{g}_0^{R,A} (\epsilon - \frac{\omega}{2}) \delta \hat{g}_{\omega_r,k}^{R,A} 
+ i D k_x \frac{\delta h}{2 v_f} [ \hat{g}_0^{R,A} (\epsilon - \frac{\omega}{2}) \tau_z \hat{g}_0^{R,A} (\epsilon + \frac{\omega}{2}) -\tau_z] = \nonumber \\
&\left[\left (\epsilon \pm i \Gamma \right) \tau_z + i \hat{\Delta}, \delta \hat{g}_{\omega_r,k}^{R,A}\right] - \lbrace\frac{\omega}{2} \tau_z, \delta \hat{g}_{\omega_r,k}^{R,A} \rbrace + i \left( \delta \hat{\Delta}_{\pm} \hat{g}_0^{R,A} (\epsilon +\frac{\omega}{2}) - \hat{g}_0^{R,A} (\epsilon - \frac{\omega}{2}) \delta \hat{\Delta}_{\pm} \right) - \hat{C}_{\phi}^{R,A}, \nonumber \\
&\hat{C}_\phi^{R,A} = e \delta \phi \left( \hat{g}_0^{R,A}(\epsilon + \frac{\omega}{2}) - \hat{g}_0^{R,A}(\epsilon - \frac{\omega}{2})\right),
\label{retarded_corr_solution}
\end{align}
and for the Keldysh component
\begin{align}
&- i \hbar  D k^2 \hat{g}_0^R(\epsilon - \frac{\omega}{2}) \delta \hat{g}^K_{\omega_r,k} -  i \hbar D k^2 \hat{g}_0^K (\epsilon - \frac{\omega}{2}) \delta \hat{g}_{\omega_r,k} ^A + i D k_x  \frac{\delta h}{2 v_f} \hat{g}_m^-
= \nonumber \\
&\left[\epsilon \tau_z + i  \hat{\Delta}, \delta \hat{g}^K_{\omega_r,k}\right] - \lbrace \left(\frac{\omega}{2}  - i \Gamma \right)\tau_z, \delta \hat{g}^K_{\omega_r,k} \rbrace + i \left( \delta \hat{\Delta}_{\pm} \hat{g}_0^K (\epsilon + \frac{\omega}{2}) - \hat{g}_0^K (\epsilon - \frac{\omega}{2}) \delta \hat{\Delta}_{\pm} \right)   - \hat{C}_\phi^K + \hat{C}_\Gamma , \nonumber \\
&\hat{g}_m^{\pm} = \hat{g}_0^K (\epsilon \pm \frac{\omega}{2}) \tau_z \hat{g}_0^A (\epsilon \mp \frac{\omega}{2}) + \hat{g}_0^R (\epsilon \pm \frac{\omega}{2}) \tau_z \hat{g}_0^K (\epsilon \mp \frac{\omega}{2}),\nonumber \\ &\hat{C}_\phi = e \delta \phi \left( \hat{g}_0^K(\epsilon + \frac{\omega}{2}) - \hat{g}_0^K(\epsilon - \frac{\omega}{2})\right), \quad \hat{C}_\Gamma = 2 i \Gamma \left[\tau_z \tanh(\frac{\epsilon - \omega/2}{2T}) \delta \hat{g}^A - \delta \hat{g}^R \tau_z \tanh(\frac{\epsilon + \omega/2}{2T}) \right].
\label{keldysh_corr_solution}
\end{align}
\end{widetext}
The second necessary component $\delta \hat g_{-\omega_r,-\bm k}$ is obtained from Eqs.~(\ref{retarded_corr_solution})-(\ref{keldysh_corr_solution}) by the substitution $(\omega, \bm k) \rightarrow (- \omega^*, - \bm k)$. It is seen that the superconducting response to the magnon is linear with respect to $\delta \bm h$ if the magnon has a non-zero component of the wave vector $k_x$  along the direction of the equilibrium exchange field. 

Solving the system of linear equations (\ref{retarded_corr_solution})-(\ref{keldysh_corr_solution}), we derive the corrections to the bulk solutions due to the perturbations of the order parameter, the exchange field and the electrical potential caused by an excitation. For the retarded and advanced components of the correction to the Green's function we obtain the following expressions:
\begin{align}
    [\delta \hat{g}_{\omega_r,k}^{R,A}]_{ij} =  a_{ij}^{R,A} \delta \Delta_{\omega_r, k} + b_{ij}^{R,A} \delta \Delta_{-\omega_r, -k}^* \nonumber \\
    + c_{ij}^{R,A} (\hat {\bm k} \times \delta \bm h_{\omega_r, \bm k}) \cdot\hat z + d_{ij}^{R,A} \delta\phi_{\omega_r, k} ,
    \label{corrections_RA}
\end{align}
where  $\hat {\bm k} = \bm k/|k|$ and the coefficients $a_{ij}^{R,A}$, $b_{ij}^{R,A}$, $c_{ij}^{R,A}$ and $d_{ij}^{R,A}$ are found from Eq.~(\ref{retarded_corr_solution}) and expressed via the equilibrium Green's functions. The explicit expressions are quite long and for this reason are provided in the Appendix.

The first-order corrections to the Keldysh Green's function are also expanded in the same manner: 
\begin{align}
        [\delta \hat{g}_{\omega_r,k}^{K}]_{ij} =  a_{ij}^{K} \delta \Delta_{\omega_r, k} + b_{ij}^{K} \delta \Delta_{-\omega_r, -k}^* \nonumber \\
        + c_{ij}^{K} (\hat {\bm k} \times \delta \bm h_{\omega_r, \bm k}) \cdot\hat z  + d_{ij}^{K} \delta\phi_{\omega_r, k},
        \label{corrections_K}
\end{align}
where the coefficients $a_{ij}^{K}$, $b_{ij}^{K}$, $c_{ij}^{K}$ and $d_{ij}^{K}$ are obtained from Eq.~(\ref{keldysh_corr_solution}) and are discussed in the Appendix.

Finally, the self-consistency equation for $\delta \Delta$ reads
\begin{equation}
    \delta \Delta_{\omega_r, k} = \frac{\lambda}{4 i} \int_{-\epsilon_c}^{\epsilon_c} d\epsilon [\delta f_{\omega_r,k}^{K}]_{12}.
    \label{self_consistency}
\end{equation}
Here $\lambda^{-1} = \int_{0}^{\epsilon_c} d \epsilon   \tanh{(\epsilon/2 T_c)}/{\epsilon} $ is the coupling constant and $\epsilon_c$ is the Debye frequency cutoff. 
The self-consistency equation for the bulk order parameter can be written as
\begin{equation}
    \Delta =\lambda \int_{0}^{\epsilon_c} d \epsilon \Re\left[\frac{\Delta}{\sqrt{(\epsilon + i\Gamma)^2 -\Delta^2}}  \tanh{\frac{\epsilon}{2 T}}\right].
\end{equation}

\subsection{Self-consistent calculation of the electric potential}
\label{subsec:scalar}
Our theory takes into account both collective modes of the superconducting order parameter: the amplitude Higgs mode and the phase NG mode. In neutral systems the phase mode is gapless with a linear dispersion law $\omega_k \propto k$, although it is well-known that in 3D superconductors the phase mode obtains a mass term and it is lifted up to the plasma frequency due to the screening by the Coulomb interaction\cite{Anderson1958_2,Anderson1963}. Here we are dealing with a 2D superconducting surface state. In 2D systems the plasmon remains a soft mode with a dispersion law $\omega_k \propto \sqrt k$, and, therefore, in superconductors the NG mode also remains soft even if the Coulomb interaction effects are taken into account \cite{ohashi1998, Buisson1994,Kliewer1967,Sun2020,lu2020,Mishonov1990}. However, the coupling of this mode to the plasmon is still important even in the 2D case,  which is confirmed by the change in the dispersion law of this mode when taking into account the Coulomb effects. Therefore, in our theory we take into account the perturbation of the electric potential $\delta \phi$, which is caused by the perturbation of the electron density $\delta n$ associated with the collective excitation. This is done by supplementing the Usadel equation with the Poisson equation for scalar potential $\phi$
\begin{equation}\label{Poisson_1}
    {\nabla}^2 \phi = -  4 \pi \rho, 
    \end{equation}
where $\rho = n e$ is the electron charge density. 

The electron density $n$ can be expressed via the Green's functions as follows\cite{Serene1983}
\begin{equation}
n = - \nu e \phi - \int \frac{d\Omega}{4 \pi} \nu \int \frac{d\epsilon}{8}  Tr [\check {\check g}(\bm n_F, \bm r, \varepsilon,t)^K],
\end{equation}
where $\nu$ is the single particle density of states at the Fermi level and the first term corresponds to the contribution due to static polarizability of the conduction band \cite{kamenev2011}. Taking into account the spin structure of the Green's function \eqref{spin_structure} we obtain
\begin{equation}\label{Poisson_1}
     {\nabla}^2 \phi = 4 \pi e^2 \nu f, \quad  f = \phi + \frac{1}{8 e} \int d\epsilon  Tr[\hat{g}^K].
\end{equation}
In 2D case which corresponds to the TS surface  the Poisson equation takes the following form
\begin{equation} \label{Poisson_2D}
    \left(\frac{\partial^2}{\partial z^2} - k^2\right) \phi_{\omega_r,k}(z) = 4 \pi e^2 \nu f_{\omega_r, k } \delta(z),
\end{equation}
where we substituted the solution of Eq.\eqref{Poisson_1} as 
\begin{equation}
    \phi(\bm r, t) = \phi_{\omega_r, k }(z) e^{i (\omega t + \bm k \cdot \bm r) }.
\end{equation}
The solution of Eq.\eqref{Poisson_2D} can be written in the form
\begin{eqnarray}
    \phi_{\omega_r, \bm k}(z) = \begin{cases}
        C e^{- k z}, z>0 \\
        C e^{ k z}, z<0 \\
    \end{cases},
\end{eqnarray}
where $C$ is to be found from the boundary condition at $z=0$: $\partial \phi_{\omega_r, k}/\partial z|_{z=+0} - \partial \phi_{\omega_r, k}/\partial z|_{z=-0} = 4 \pi e^2 \nu f_{\omega_r, k}$. Then the 2D scalar potential $\phi_{\omega_r, k}(z=0) \equiv \phi_{\omega_r, k}$ takes the form: 
\begin{equation}\label{Poisson_final_2D}
    \phi_{\omega_r, k} =-\frac{\hbar^2\omega_p^2}{4\pi (k \xi)(\pi k_B T_c)\Delta_0 } f_{\omega_r, k},
\end{equation}
where $\omega_p = \sqrt{ 4 \pi^2 e^2 \nu \Delta_0 D/ \hbar\xi}$ is the superconducting plasma frequency  and $\Delta_0= \Delta (T=0, \Gamma\rightarrow0)$ \cite{kamenev2011,Sun2020}.

Using Eq.~(\ref{Poisson_1}) we can express scalar potential $\delta\phi_{\omega_r, k}$ in terms of the excitation correction to the Keldysh Green's function. The scalar potential can be written as
\begin{align}
    \delta\phi_{\omega_r, k} = - \frac{\tilde{\omega}_p^2}{(1 + n_{\delta\phi}) \tilde{\omega}_p^2 + 2 k \xi} \left[n_{\delta \Delta} \delta\Delta_{\omega_r, k} \right. \nonumber \\
    \left. +n_{\delta \Delta^*} \delta\Delta_{-\omega_r,- k}^* + n_{\delta h} (\hat {\bm k} \times \delta \bm h_{\omega_r, \bm k}) \cdot\hat z \right],
\end{align}
where
\begin{align}
    &n_{\delta \Delta} = \frac{1}{8 e} \int d \epsilon \left[a_{11}^K +a_{22}^K\right], \\
    &n_{\delta \Delta^*} = \frac{1}{8 e} \int d \epsilon \left[b_{11}^K +b_{22}^K\right], \\
    &n_{\delta h} = \frac{1}{8 e} \int d \epsilon \left[c_{11}^K +c_{22}^K\right],\\
    &n_{\delta \phi} = \frac{1}{8 e} \int d \epsilon \left[d_{11}^K +d_{22}^K\right],
\end{align}
and $\tilde{\omega}_p = \hbar \omega_p /\sqrt{ 2 \pi \Delta_{0}\pi k_B T_c}$.
Substituting the scalar potential into Eqs.~\eqref{corrections_RA} and \eqref{corrections_K} we can obtain the Green's function corrections with embedded Poisson equation
\begin{align}\label{dg_K_poisson}
        [\delta \hat{g}_{\omega_r,k}^{R,A,K}]_{ij} =  &\Aa_{ij}^{R,A,K} \delta \Delta_{\omega_r, k} + \Ba_{ij}^{R,A,K} \delta \Delta_{-\omega_r, -k}^* \nonumber \\
        &+ \Ca_{ij}^{R,A,K} (\hat {\bm k} \times \delta \bm h_{\omega_r, \bm k}) \cdot\hat z,
\end{align}
where
\begin{eqnarray}\label{coeffs_Poisson}
    \Aa_{ij}^{R,A,K} = a_{ij}^{R,A,K} -  \frac{\tilde{\omega}_p^2 n_{\delta \Delta}}{(1 + n_{\delta\phi}) \tilde{\omega}_p^2 + 2 k \xi}d_{ij}^{R,A,K}, \\
    \Ba_{ij}^{R,A,K} = b_{ij}^{R,A,K} -  \frac{\tilde{\omega}_p^2 n_{\delta \Delta^*}}{(1 + n_{\delta\phi}) \tilde{\omega}_p^2 + 2 k \xi}d_{ij}^{R,A,K}, \\
    \Ca_{ij}^{R,A,K} = c_{ij}^{R,A,K} -  \frac{\tilde{\omega}_p^2 n_{\delta h}}{(1 + n_{\delta\phi}) \tilde{\omega}_p^2 + 2 k \xi}d_{ij}^{R,A,K}.
\end{eqnarray}
The excitation-induced correction to the Keldysh Green's function expressed by Eq.~(\ref{dg_K_poisson}) is the main quantity describing the electronic part of collective excitations of hybrid systems consisting of a charged superconductor and a magnetic insulator.

\subsection{Magnetic part of the hybrid collective excitations}
\label{subsec:magnon}
The dynamics of the spin wave in the ferromagnetic insulator is described by the LLG equation (\ref{LLG}).
%
The electron spin polarization $\bm s$ in the helical metal can be calculated via the electric current in the TS  according to Eq.~(\ref{polarization_current}).

In order to calculate the electrical current we use the following relation \cite{Bobkova2016} 
\begin{equation}
    \bm J = \frac{\sigma}{16 e} \int d\epsilon Tr \left[\tau_z \check{g} \otimes \bm \hat{\nabla} \check{g}\right]^K,
    \label{eq:current_general}
\end{equation}
where $\sigma$ is the conductivity of the TS conductive surface state. According to Eq.~(\ref{delta_correction}) the electric current also takes the form: 
\begin{equation}
    \bm J = \bm J_{\omega_r,k} e^{i \left(\omega t + \bm k \cdot \bm r \right)} +  \bm J_{-\omega_r,-k} e^{-i \left(\omega^* t + \bm k \cdot \bm r \right)},
\end{equation}
where
\begin{align}
&\bm J_{\omega_r,k} =  \frac{\sigma}{16 e} \int d \epsilon \Bigl\{ i \bm k \left(g_0^K(\epsilon - \frac{\omega}{2}) (\delta g^A_{11} + \delta g^A_{22}) \right.\nonumber \\
&+ g_0^R(\epsilon - \frac{\omega}{2}) (\delta g^K_{11} + \delta g^K_{22})+  
 i f_0^K(\epsilon - \frac{\omega}{2}) ( \delta f^A_{21} - \delta f^A_{12}) \nonumber \\
 &\left. + i f_0^R(\epsilon - \frac{\omega}{2}) ( \delta f^K_{21} - \delta f^K_{12}) \right)   - i  {\hat{x}} \frac{\delta h}{\hbar v_f} \Gamma_{xx}^K\Bigr\}, 
\label{current_corrections}
\end{align}
Since $\bm m_0$ is along the $x$-axis, the only component of the excitation-induced electric current entering the torque term in Eq.~(\ref{LLG}) is $J_x$. It also can be expanded over the different components of the composite excitation:
\begin{eqnarray}
J_{\omega_r, \bm k}^x  = (\hat {\bm k}\cdot\hat{x})\left[J_{\delta \Delta} \delta \Delta_{\omega_r, \bm k} + J_{\delta \Delta^*} \delta \Delta^*_{-\omega_r,- \bm k}  \right. \nonumber \\
\left. + J_{2,(\omega_r, \bm k)}[\hat {\bm k} \times \delta \bm h _{\omega_r, \bm k}]\cdot\hat z\right] +  J_{0,\omega_r} (\delta \bm h_{\omega_r, \bm k} \times \hat z) \cdot\hat{x}.
\label{current_expansion}
\end{eqnarray}
Coefficients $J_{\delta \Delta}$, $J_{\delta \Delta^*}$, $J_{0,\omega_r}$ and $J_{2,(\omega_r, \bm k)}$ are found from  Eqs.~(\ref{current_corrections}) and (\ref{dg_K_poisson}).

\subsection{Calculation of the spectrum of the hybrid collective modes}
\label{subsec:total_spect}
The hybrid excitations, which consist of the superconducting order parameter excitations in the TS and magnons in FI, are investigated in the basis $\hat \Psi = (\delta \Delta^a, \delta \Delta^p, \delta h_y, \delta h_z)^T$. Here the first two components $\delta \Delta^{a}_{\omega_r, \bm k} =[\delta \Delta_{\omega_r, \bm k} + \delta \Delta_{-\omega_r, -\bm k}^* ]/2$ and $\delta \Delta^{p}_{\omega_r, \bm k} =[\delta \Delta_{\omega_r, \bm k} - \delta \Delta_{-\omega_r, -\bm k}^* ]/2i$ represent the amplitude and phase modes of the superconducting order parameter, respectively. The second two components describe the magnetic part of the excitation, that is the magnon, via the relation $\delta h_{y,z} = -J_{ex} M_s \delta m_{y,z} / (2 \gamma d_s)$. The  spectrum of the hybridized excitations is found from the combination of the self-consistency equation (\ref{self_consistency}) for the OP and the linearized with respect to $\delta \bm m$ and $\bm s$ LLG equation (\ref{LLG}). The resulting  linear matrix equation for finding the eigenmodes of the FI/TS system can be written in the form:
\begin{equation}
    \begin{pmatrix}
        \hat M_{\Delta \Delta} & \hat M_{\Delta h} \\
        \hat M_{h \Delta } & \hat M_{h h}
    \end{pmatrix} \hat \Psi =0,
    \label{matrix_modes}
\end{equation}
where $\hat M_{\Delta \Delta}$, $\hat M_{\Delta h}$, $\hat M_{h \Delta}$ and $\hat M_{hh}$ are $2\times 2$ matrices depending on $(\omega, \bm k)$.  The first two lines of Eq.~(\ref{matrix_modes}) are nothing but the self-consistency equation (\ref{self_consistency}) and its complex-conjugate. $\hat M_{\Delta \Delta}$ takes the form
\begin{widetext}
\begin{eqnarray}
\hat M_{\Delta \Delta} = 
\left(
\begin{array}{cc}
A_{12}^K(\omega, \bm k) - 1 + B_{12}^K(\omega, \bm k)  & i\left[ A_{12}^K(\omega, \bm k) - 1 - B_{12}^K(\omega, \bm k) \right] \\
A_{12}^{K*}(-\omega^*, -\bm k) - 1 + B_{12}^{K *}(-\omega^*, -\bm k) &  -i \left [A_{12}^{K*}(-\omega^*, -\bm k) - 1 - B_{12}^{K *}(-\omega^*, -\bm k) \right]
\end{array}
\right)
\label{M_delta_delta},
\end{eqnarray}
\end{widetext}
where 
\begin{eqnarray} \label{ab_integrated}
    A_{12}^K (\omega, \bm k)= \frac{\lambda}{4 i} \int_{-\epsilon_c}^{\epsilon_c} d\epsilon \Aa_{12}^K(\epsilon, \omega, \bm k), \nonumber \\
    B_{12}^K (\omega, \bm k)= \frac{\lambda}{4 i} \int_{-\epsilon_c}^{\epsilon_c} d\epsilon \Ba_{12}^K(\epsilon, \omega, \bm k).
\end{eqnarray}
Matrix $\hat M_{\Delta h}$ accounts for the possibility of the dynamic OP corrections excited by the linear coupling to magnons and takes the form:
\begin{eqnarray}
\hat M_{\Delta h} = 
\left(
\begin{array}{cc}
sgn(k_x)C_{12}^K(\omega, \bm k)  & 0 \\
-sgn(k_x)C^{K*}_{12}(-\omega^*,- \bm k)   & 0 
\end{array}
\right)
\label{M_delta_h},
\end{eqnarray}
where 
\begin{equation}\label{c_integrated}
      C_{12}^K (\omega, \bm k)= \frac{\lambda}{4 i} \int_{-\epsilon_c}^{\epsilon_c} d\epsilon \Ca_{12}^K(\epsilon, \omega, \bm k).
\end{equation}
The zero second column of $\hat M_{\Delta h}$ reflects the fact that in the considered limit $\mu \gg (\Delta,h)$ only in-plane $y$-component of the magnon magnetization interacts with the TS.

The bottom  two lines of Eq.~(\ref{matrix_modes}) represent the linearized LLG equation (\ref{LLG}), where $\delta \bm m$ is expressed via $\delta \bm h$ as $\delta \bm m = -2\gamma d_s \delta \bm h/(J_{ex})M_s$. Matrices $\hat M_{h\Delta}$ and $\hat M_{hh}$ take the form:
\begin{widetext}
\begin{eqnarray}
\hat M_{h \Delta} = 
\left(
\begin{array}{cc}
0  & 0 \\
c_s {\rm sgn}(k_x) \left [ J_{\delta \Delta} (\omega, \bm k) 
 + J_{\delta \Delta^*} (\omega, \bm k) \right]  & 
 i c_s {\rm sgn}(k_x)\left [ J_{\delta \Delta} (\omega, \bm k) 
 - J_{\delta \Delta^*} (\omega, \bm k) \right]
\end{array}
\right),
\label{M_h_delta}
\end{eqnarray}
\begin{eqnarray}
\hat M_{h h} = 
\left(
\begin{array}{cc}
i \omega   & (\omega_b + i \alpha \omega) \\
-(\omega_b + i \alpha \omega) +  c_s \left[J_{0,\omega_r} + {\rm sgn}(k_x) J_{2,(\omega_r, \bm k)} \right]  & i\omega 
\end{array}
\right),
\label{M_h_h}
\end{eqnarray}
\end{widetext}
where $c_s = h_0 J_{ex} /( e v_f d_f) = { 2 h_0^2 \gamma d_s}/{e v_f M_s d_f }$, and $\omega_b = \omega_0 + D_m k^2$ with $\omega_0 = \gamma K$ is the bare magnon dispersion. 

\section{Spectrum of the hybrid collective excitations}
\label{sec:spectrum}

Now we can find the response of the superconductor to magnon induced effective exchange field $\delta h$. Substituting Eq.~(\ref{dg_K_poisson}) into the self-consistency equation (\ref{self_consistency}) we obtain
\begin{equation}
    \delta \Delta_{\omega_r, \bm k}^{p,a} = F_{\omega_r,k}^{p(a)}(\hat {\bm k} \times \delta \bm h_{\omega_r, \bm k}) \cdot\hat z,
\end{equation}
with
\begin{widetext}
    \begin{align}
    F_{\omega_r,\bm k}^a &=\frac{-C^{K}_{12} (\omega, \bm k) \left(A_{12}^{K*} (-\omega^*, -\bm k) - 1 - B_{12}^{K*} (-\omega^*, -\bm k)\right)+ C^{K*}_{12} (-\omega^*, -\bm k) \left(A_{12}^{K} (\omega, \bm k) - 1 - B_{12}^{K} (\omega, \bm k)\right)}{i \cdot det\left[\hat{M}_{\Delta\Delta}\right]} , \label{eq:response_Higgs} \\
    F_{\omega_r,\bm k}^p &=\frac{C^{K}_{12} (\omega, \bm k) \left(A_{12}^{K*} (-\omega^*, -\bm k) - 1 + B_{12}^{K*} (-\omega^*, -\bm k)\right)+C^{K*}_{12} (-\omega^*, -\bm k) \left(A_{12}^{K} (\omega, \bm k) - 1 + B_{12}^{K} (\omega, \bm k)\right)}{  det\left[\hat{M}_{\Delta\Delta}\right]}  ,
    \label{eq:response_NG}
\end{align}
 where
 \begin{equation}    det\left[\hat{M}_{\Delta\Delta}\right] =-2i \left[ \left (A_{12}^{K} (\omega, \bm k) - 1 \right) \left(A_{12}^{K*} (-\omega^*, -\bm k) - 1\right) - B_{12}^{K} (\omega, \bm k) B_{12}^{K*} (-\omega^*, -\bm k) \right] .
\end{equation}
\end{widetext}

\subsection{Higgs mode}
\label{sec:higgs}

In Appendix \ref{app:amplitude_response} it is shown that $ A_{12}^{K*} (-\omega^*, -\bm k) = A_{12}^{K} (\omega, \bm k)$, $ B_{12}^{K*} (-\omega^*, -\bm k) = B_{12}^{K} (\omega, \bm k)$ and $ C_{12}^{K*} (-\omega^*, -\bm k) = C_{12}^{K} (\omega, \bm k)$. Making use of these symmetry relations from Eq.~(\ref{eq:response_Higgs}) we immediately obtain that $F_{\omega_r, \bm k}^a = 0$ for the TS. It means that if the TS is in the helical ground state the amplitude Higgs mode is not coupled to the magnon in the linear order.

\begin{figure}[t]
\includegraphics[width=\columnwidth]{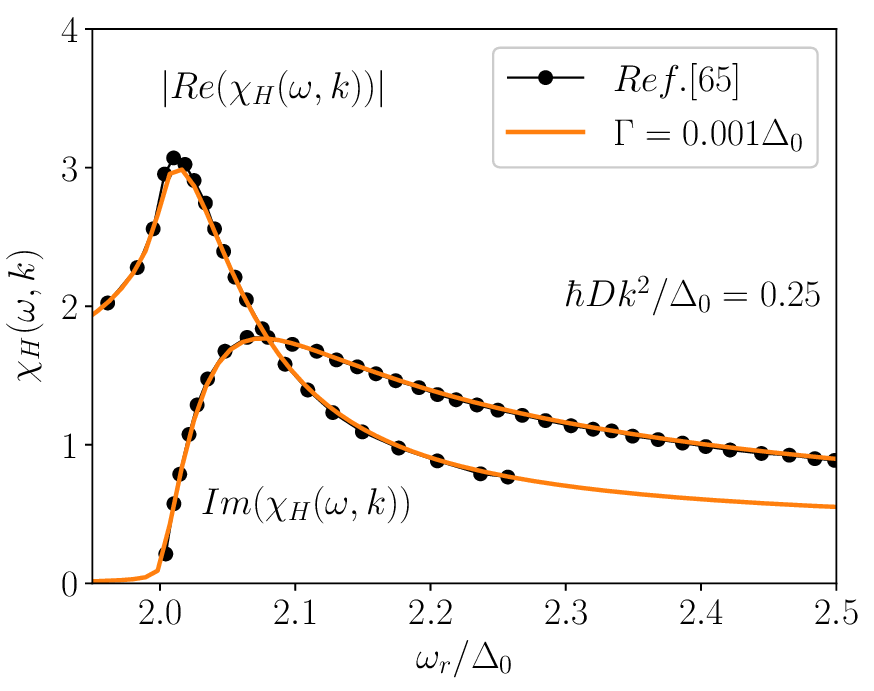}
\caption{ Real and imaginary parts of the Higgs mode susceptibility $\chi_H(\omega, k)$ as a function of the excitation frequency at $D k^2/\Delta_0 = 0.25$ and $T=0.1 T_c$ (orange lines) in comparison to the corresponding results taken from Ref.~\cite{Nosov2024} (black points).} 
 \label{fig:Higgs}
\end{figure}

Then since the diffusive TS in the helical state is fully equivalent to a conventional diffusive $s$-wave singlet superconductor, the Higgs mode in the helical state of the TS is fully equivalent to the Higgs mode of the disordered $s$-wave superconductor, which was studied in details in Ref.~\cite{Nosov2024}. In particular, it was found that the frequency of the Higgs mode can be below $2 \Delta$ for sufficiently strong disorder, while the spectral function  exhibits a wide peak above the edge of the two-particle continuum. In Fig.~\ref{fig:Higgs} we present the results for the (charge neutral) susceptibility $\chi_H= \lambda \left[A_{12}^{K} (\omega, \bm k) +B_{12}^{K} (\omega, \bm k) - 1\right] ^{-1}$, calculated in the framework of the theory presented in the previous section, which are in excellent agreement with the results of Ref.~\cite{Nosov2024}.

\subsection{NG mode in the absence of the coupling to FI}

In the absence of the exchange coupling to the FI equation $A_{12}^K(\omega, \bm k) - 1 - B_{12}^K(\omega, \bm k) = 0$ gives the phase Nambu-Goldstone mode.

\begin{figure}[t]
\includegraphics[width=0.9\columnwidth]{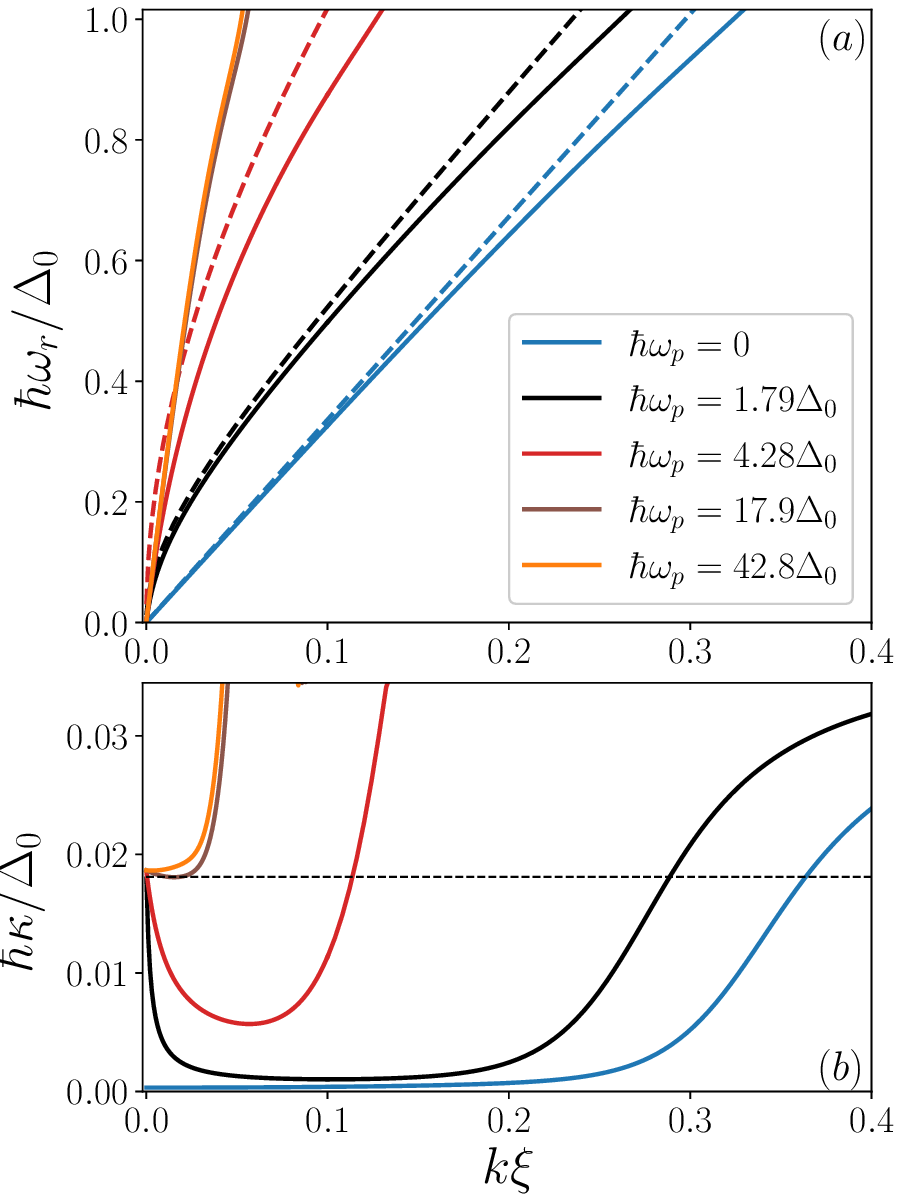}
\caption{ The spectrum (a) and decay rate (b) of superconducting phase mode at various plasma frequencies calculated at $\Gamma = 0.018 \Delta_0$ and $T=0.1 T_c$, which is shown as a horizontal dashed line. Color dashed lines in panel (a) are results of analytical calculation making use of Eq.~(\ref{NG_analytical}).}
 \label{phase_intrinsic}
\end{figure}

\begin{figure}[t]
\includegraphics[width=1\columnwidth]{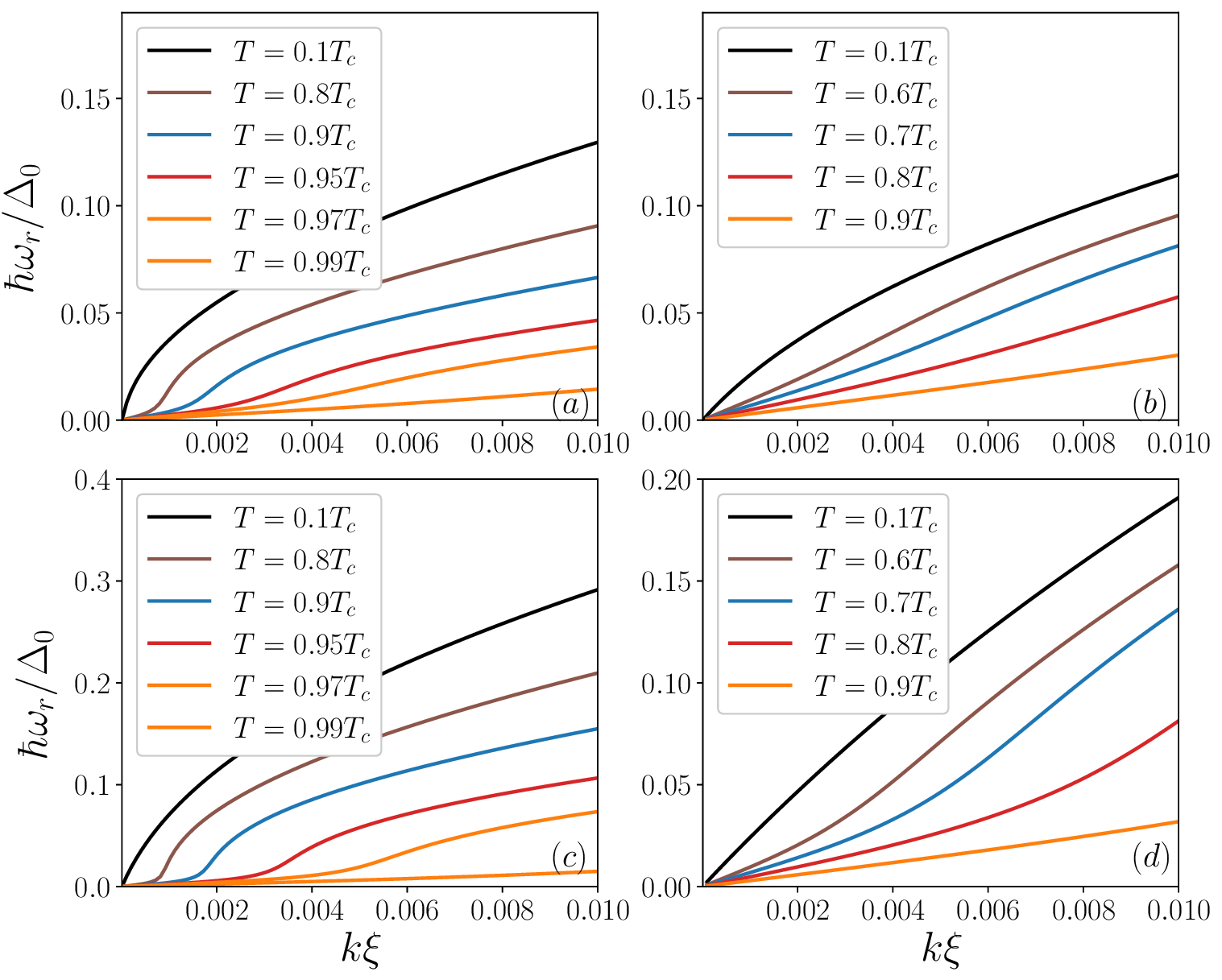}
\caption{The spectrum of superconducting phase mode at various temperatures: (a) - $\hbar\omega_p = 1.79 \Delta_0$ and $\Gamma = 0.0018 \Delta_0$, (b) - $\hbar\omega_p =  1.79\Delta_0$ and $\Gamma = 0.018 \Delta_0$, (c) - $\hbar\omega_p = 4.28 \Delta_0$ and $\Gamma = 0.0018 \Delta_0$, (d) - $\hbar\omega_p = 4.28 \Delta_0$ and $\Gamma = 0.018 \Delta_0$.}
 \label{phase_intrinsic_crossover}
\end{figure}
The dispersion $\omega(k)$ and the decay rate $\kappa (k)$ of the NG mode calculated according to this equation are shown in Fig.~\ref{phase_intrinsic} for different plasma frequencies. At low temperatures $T \ll \Delta$, low frequencies $\omega \ll \Delta$, small wavenumbers $k \xi \ll 1$ and {$\Gamma \ll \omega$}  the spectrum of the NG mode of the charged 2D superconductor can be found analytically. For this purpose we follow the procedure of Ref.\cite{kamenev2011} and arrive at the following matrix equation
\begin{eqnarray}
    \begin{pmatrix}
        \frac{(\hbar\omega_r)^2 - \pi \Delta_0  \hbar D k^2}{ 4 \Delta_0^2} & \frac{i \hbar \omega_r}{2 \Delta_0} \\
         -\frac{i \hbar \omega_r}{2 \Delta_0}&  1 + \frac{k}{2 \pi e^2 \nu}
    \end{pmatrix} \begin{pmatrix}
        \delta\Delta_p \\
        e \delta \phi_{\omega_r,k}
    \end{pmatrix}=0.~~~~
    \label{eq:NG_derivation_1}
\end{eqnarray}
which results in dispersion relation
\begin{align}
\omega^2_{NG}= 2 \Delta_0 \pi^2 k_B T_c \hbar^{-2} \left( \frac{1}{2} \tilde{\omega}_p^2 k\xi + ( k \xi)^2\right).
\label{NG_analytical}
\end{align}
The analytical expression Eq.~(\ref{NG_analytical}) is in reasonable agreement with our numerical result presented in Fig.~\ref{phase_intrinsic}(a) in the appropriate range of parameters $\omega \ll \Delta$, $k \xi \ll 1$ and {$\Gamma \ll \omega$}. The dispersion relation of the NG mode demonstrates a crossover between the linear and square root in momentum behavior with increase of the superconducting plasma frequency. We note that the collective excitations of the phase in dirty superconductors were extensively studied in Ref.\cite{Karuzin2025}.

In Fig.\ref{phase_intrinsic_crossover} we show results for the NG mode calculated at wide temperature range.
It is seen that at low temperature and small $\Gamma$ this mode is close to the square-root behavior (Fig. \ref{phase_intrinsic_crossover}(a) and (c), black curves), which indicates the NG-plasmon coupling. At the same time, at higher temperatures or at higher values of $\Gamma$ (see other curves in Fig.~\ref{phase_intrinsic_crossover}) the behavior gradually evolves from square-root to linear, which means that the NG mode becomes uncoupled from the plasmon. This is due to the presence of thermally-induced quasiparticles at high temperatures or even at low temperatures if the gap is smeared by substantial $\Gamma$. Then the supercurrents induced by the phase mode are counterbalanced by the normal currents from thermally excited quasiparticles, without any need to generate electron density oscillations. In this case the gapless phase excitation is called the Carlson-Goldman mode and has a linear dispersion law \cite{Carlson1975,Schmid1975, Artemenko1979,Karuzin2025}.

\subsection{Hybridized magnon-NG mode}
\label{subsec:magnonNG}
\begin{figure}[t]
\includegraphics[width=85mm]{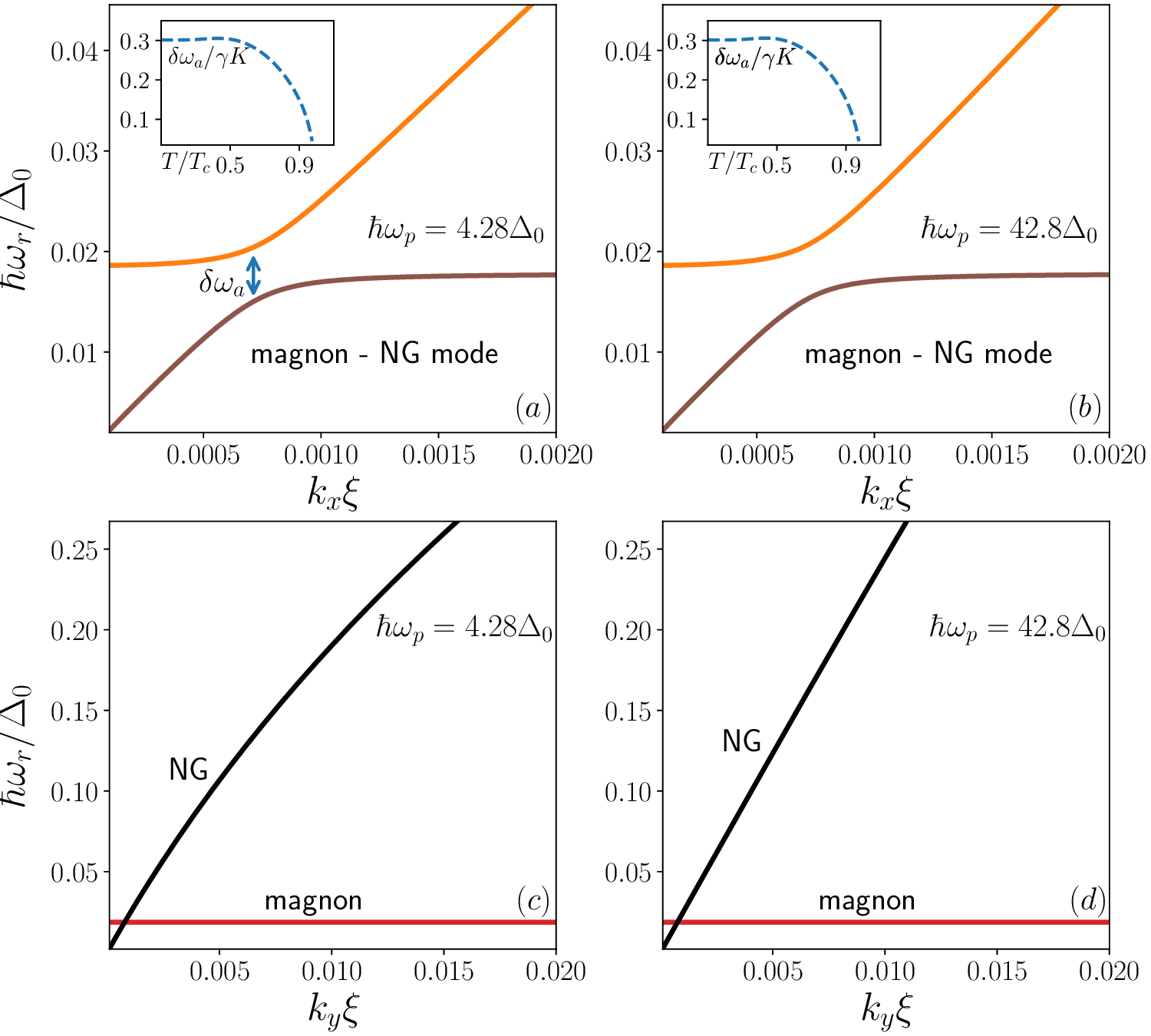}
\caption{Hybridization between NG and magnon modes. Left column: $\hbar\omega_p = 4.28 \Delta_0$; right column - $\hbar\omega_p = 42.8\Delta_0$. Top row: excitations propagating along the $x$-axis; bottom row: along the $y$-axis.
The anticrossing strength $\delta \omega_a$ is defined in Eq.~\eqref{domega_a}. The parameters used for the calculations are as follows.
Superconducting subsystem: $\Gamma = 0.018 \Delta_0$, $T=0.1 T_c$, $\xi =3.5 nm$ \cite{Onar2016} and $\sigma = 3 \cdot 10^{14} c^{-1}$ (normal state conductivity).
Magnetic subsystem: $\gamma = 1.76 \cdot 10^7 G^{-1} s^{-1}$, $\omega_0 = \gamma K = 10^{-17} erg \approx 0.018 \Delta_0 $, $D_m=5 \cdot 10^{-29} erg \cdot cm^2$ \cite{Xiao2010} and $  4 \pi M_s \approx  2 \cdot 10^3 Oe$ \cite{kajiwara2010}.
The exchange field is set to $ h_0 \approx 4.16 \Delta_0$.
Insets in panels (a) and (b) display the anticrossing strength $\delta \omega_a$ as a function of the temperature. 
}
 \label{hybr}
\end{figure}

As we have shown in Sec.~\ref{sec:higgs} the system does not support amplitude response to the magnon excitation. Instead the superconducting subsystem responds well to the magnon in the form of transverse oscillations, i. e. phase oscillations $\delta \Delta^p$. For this reason we reduce the basis vector $\hat \Psi \to \hat \Psi^p = (\delta \Delta^p, \delta h_y)^T$.  Then the spectrum of the collective excitations can be calculated from the following matrix equation

\begin{eqnarray}
    \begin{pmatrix}
        M^p & {\rm sgn}k_x C_{12}^K(\omega, \bm k) \\
        c_s {\rm sgn}(k_x)J_p &  M^h
    \end{pmatrix} \hat \Psi_p =0,
    \label{eq:dispersion_eq_reduced}
\end{eqnarray}
where $M^p = i \left[ A_{12}^K(\omega, \bm k) - 1 - B_{12}^K(\omega, \bm k) \right]$, $J_p = i \left [ J_{\delta \Delta} (\omega, \bm k) 
 - J_{\delta \Delta^*} (\omega_, \bm k) \right] $ and $M^h = \omega^2/(\omega_b + i \alpha \omega)-(\omega_b + i \alpha \omega)+ c_s [J_{0,\omega_r} + {\rm sgn}(k_x) J_{2,(\omega_r, \bm k)}]$.

The dispersion curves of the hybridized magnon-NG mode were calculated numerically from Eq.~(\ref{eq:dispersion_eq_reduced}) are shown here in Fig.~\ref{hybr}. First of all, due to the spin-momentum locking which dictates the symmetry of the linear response of the OP on the magnon in the form $(\hat {\bm k} \times \delta \bm h_{\omega_r, \bm k}) \cdot \hat z$ the coupling is anisotropic. Its magnitude is maximal if the magnon propagates along the equilibrium magnetization direction ($x$-axis), leading to the anticrossing between the NG and the magnon modes, and is zero if the magnon propagates along the $y$-axis, as it is demonstrated in Fig.~\ref{hybr}. Here we focus on the analytical description of spectrum in the vicinity of the anticrossing region for the case of maximal hybridization $\hat {\bm k} = \hat x$.

At the intersection of bare magnon $\omega_b = \omega_0 + D_m k^2$ and NG  mode [Eq.~(\ref{NG_analytical})] dispersions, we can find that the momentum of the intersection point takes the form
\begin{align}
    k_{i} \xi = \frac{2 \Delta \pi k_B T_c \tilde{\omega}_p^2  - \sqrt{(2 \Delta \pi^2 k_B T_c \tilde{\omega}_p^2)^2 -8 \omega_0^3 D_m \xi^{-2}}}{4 \omega_0 D_m \xi^{-2}}.
\end{align}
For normal state conductivity $\sigma = 3 \cdot 10^{14} c^{-1} $, $d_s = \xi$, $\xi = 3.5 nm$ \cite{Onar2016} and $T_c = 14.5 K$ which yields $\Delta_0/ \hbar \approx 3.34 THz$, we get the following estimation of 2D plasma frequency 
\begin{align*}
   \omega_p &=   \sqrt{\frac{4 \pi^2 e^2 \nu \Delta_0 D}{\hbar\xi}}  \nonumber \\  &= \sqrt{\frac{2 \pi^2 \sigma d_s \Delta_0}{ \hbar\xi}} \approx  142  THz \approx 43 \Delta_0/\hbar.
\end{align*}
This result suggests that from the experimental point of view the most relevant limit is $\tilde{\omega}_p \gg1$. In this limit we obtain
\begin{align}\label{k_anti_large_pl}
    k_{i} \xi \approx \frac{\omega_0^2}{2 \Delta \pi^2 k_BT_c \tilde{\omega}_p^2}.
\end{align}
Assuming magnetic parameters corresponding to YIG $\gamma = 1.76 \cdot 10^7 G^{-1} s^{-1}$, $\omega_0 = \gamma K = 10^{-17} erg \approx 0.018 \Delta_0 $, $  4 \pi M_s \approx  2 \cdot 10^3 Oe$ \cite{kajiwara2010} , and $D_m=5 \cdot 10^{-29} erg \cdot cm^2$ \cite{Xiao2010}, we obtain that the hybridization region corresponds to the limit $k_i \xi \ll 1$, as it is also seen in Fig.~\ref{hybr}. 

In the limit of low temperature, $k\xi\ll 1$ and $\omega\ll\Delta$ and using the limiting expressions for the self-consistency equations reported in Ref.\cite{kamenev2011}~, Eq.~(\ref{eq:dispersion_eq_reduced}) can be written as follows
\begin{widetext}
\begin{eqnarray}
    \begin{pmatrix}
        \hbar^2|k_x \xi|(\omega^2 -\omega_{NG}^2)/4 \Delta_0^2 x_c \left[k_x\xi + \tilde{\omega}_p^2\right]   &  {\rm sgn}(k_x)  C_{12}^K \\
        c_s {\rm sgn}(k_x)J_p &  \left[\omega^2 - \omega_b^2\right]/\omega_b + c_s(J_{0} + J_{2})
    \end{pmatrix} \hat \Psi_p =0,
\end{eqnarray}
\begin{align}
    C_{12}^K \approx  - \frac{\lambda \pi k_B T_c}{4 i} \int_{-\epsilon_c}^{\epsilon_c}d\epsilon \frac{ |k_x \xi^2|\Gamma_{xy}^K(\omega =0) }{2 \hbar v_f \epsilon}, \\
    J_p \approx  \frac{2 i|k_x \xi| \sigma}{16 e \xi}\int_{-\epsilon_c}^{\epsilon_c}d\epsilon \left( \frac{f_0^R(\epsilon) g_0^K(\epsilon)}{\epsilon} + \frac{f_0^K(\epsilon) g_0^A(\epsilon)}{\epsilon - i\Gamma}- \frac{2 i \Gamma \Delta_0 (g_0^R(\epsilon))^2 \tanh{(\epsilon/2T)}}{\epsilon (\epsilon + i \Gamma) (\omega - 2i \Gamma)} \right), \\
    J_0 \approx - i \frac{\sigma}{16 e} \int_{-\epsilon_c}^{\epsilon_c}d\epsilon \frac{\Gamma_{xx}^K(\omega=0)}{\hbar v_f}, \\
    J_2 \approx - \frac{\sigma}{16e} \int_{-\epsilon_c}^{\epsilon_c}d\epsilon \frac{ \pi k_B T_c(k_x \xi)^2}{\hbar v_f} \left( \frac{f_0^K(\epsilon) \Gamma_{xy}^A(\omega =0)}{\epsilon - i \Gamma} + \frac{f_0^R(\epsilon) \Gamma_{xy}^K(\omega =0)}{\epsilon}\right),
\end{align}
\end{widetext}
where $x_c \approx 4.27$, Keldysh coefficients $\Gamma_{xx}^K$ and $\Gamma_{xy}^K$ are defined by Eqs.~(\ref{eq:Gamma_xx_K}), (\ref{eq:Gamma_xx_K}), respectively. Assuming $\Gamma\ll\omega$, we neglect the last term in $J_p$, and the spectrum of the collective excitations can be found from the following equation
\begin{align}\label{spec_simple}
    \left(\omega^2 -\omega_{NG}^2\right) \left(\left[\omega^2 - \omega_b^2 \right]/\omega_b + c_s (J_0 + J_2)\right) \nonumber \\ -  4\beta c_s \left(\tilde{\omega}_{p}^2 k_x \xi\ + (k_x \xi)^2\right) = 0.
\end{align}
Here $\beta = \pi k_B T_c x_c \xi \Delta_0^2 \tilde{J}_p \tilde{C}_{12}^K$, where $\tilde{C}_{12}^K = C_{12}^K/ \pi k_B T_c |k_x \xi^2|$ and $\tilde{J}_p = J_p /|k_x \xi|$. Using the approximation
\begin{align}
    \left(\omega^2 -\omega_{NG}^2\right) \left(\left[\omega^2 - \omega_b^2 \right]/\omega_b + c_s (J_0 + J_2)\right) \nonumber \\ \approx  4 \omega_{NG}\left(\omega -\omega_{NG}\right) \left(\omega - \omega_b'\right),
    \label{eq:approx_analytical_1}
\end{align}
where $\omega_b' = \omega_b - c_s (J_0 + J_2)/2$ is the renormalized magnon spectrum. As it was demonstrated in Ref.~\cite{Bobkova2022}, the main effect of the renormalization at gigahertz frequencies $\omega \ll \Delta$ is  the renormalization of the magnon stiffness $D_m$. The renormalization of the zero-momentum magnon frequency $\omega_0$ is negligible. However, as it was discussed above, in the considered case the magnon dispersion is "flat" and $\omega_b' \approx \omega_b \approx \omega_0$. Substituting Eq.~(\ref{eq:approx_analytical_1}) into Eq.~\eqref{spec_simple} we obtain
\begin{align}
    \omega^2 - (\omega_b' + \omega_{NG}) \omega  + \omega_b'\omega_{NG} - \nonumber \\
    \frac{\beta c_s}{\omega_{NG}} \left(\tilde{\omega}_{p}^2 k_x \xi\ + (k_x \xi)^2\right)=0.
    \label{eq:equation_NG_magnon_analytical}
\end{align}
The solution of Eq.~(\ref{eq:equation_NG_magnon_analytical}) takes the form
\begin{align}
    &\omega_{up,dn} = \frac{1}{2} \left(\omega_b' + \omega_{NG} \pm \right. \nonumber \\
    &\left. \sqrt{\left(\omega_b' - \omega_{NG}\right)^2 + 4 \frac{\beta c_s}{\omega_{NG}}\left(\tilde{\omega}_{p}^2 k_x \xi\ + (k_x \xi)^2\right) }\right).
    \label{eq:NG_magnon_analytical_result}
\end{align}
\begin{figure}[t]
\includegraphics[width=\columnwidth]{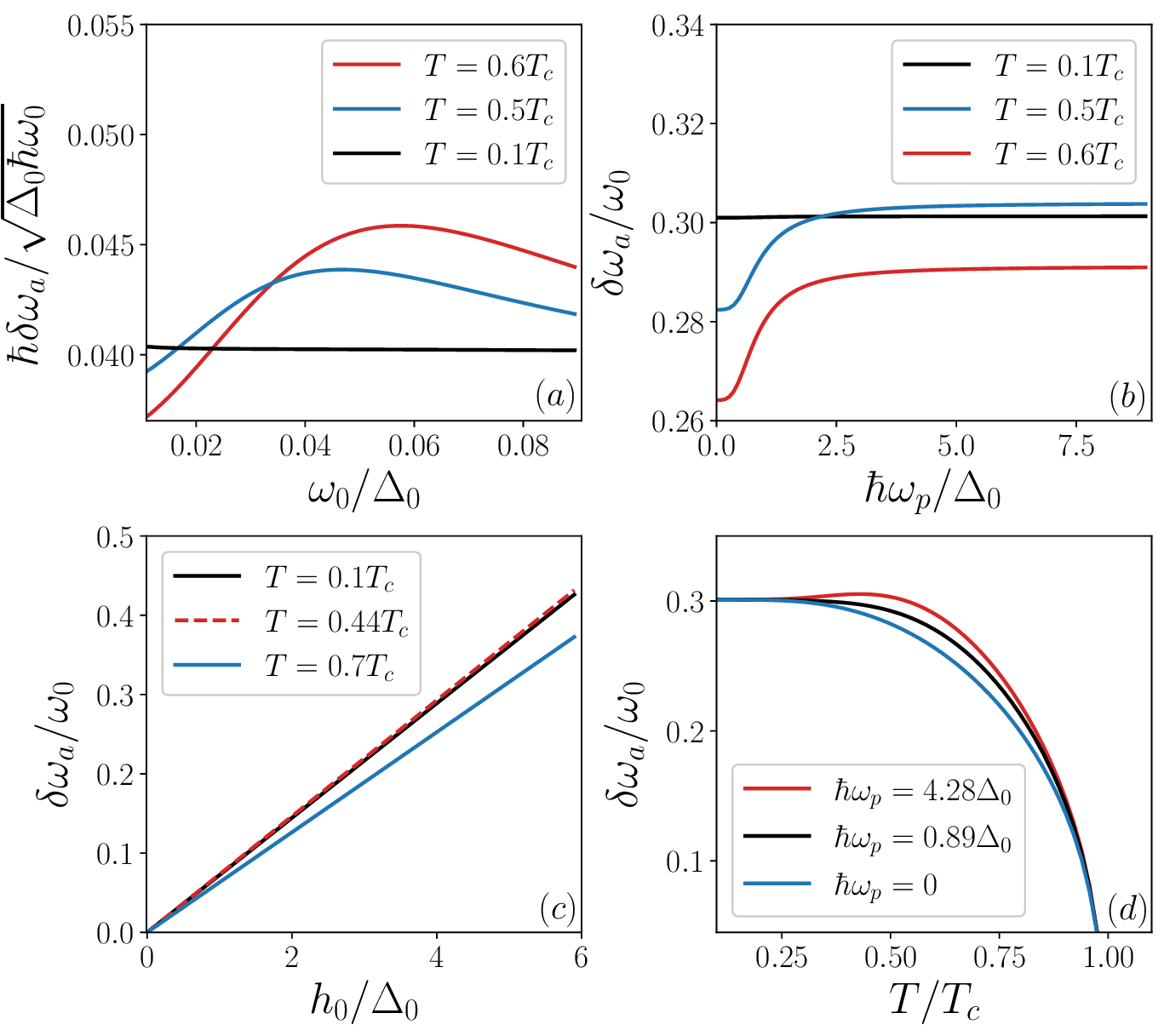}
\caption{Anticrossing strength $\delta \omega_a =\omega_{up}(k_i) - \omega_{dn}(k_i)$ as a function of $\omega_0$ (a), $\omega_p$ (b), effective exchange field $h_0$ (c) and temperature $T$ (d). In plots (a) and (c) $
\omega_p = 4.28 \Delta_0$, while $h_0 \approx 4.16 \Delta_0$ in plots (a), (b) and (d). All other parameters are consistent with Fig.\ref{hybr}. } 
 \label{anti_xi}
\end{figure}
Then we examine the anticrossing strength $\delta \omega_a =\omega_{up}(k_i) - \omega_{dn}(k_i)$ as a function of relevant physical parameters. Since $\omega_b'(k_i)  = \omega_{NG}(k_i) \approx \omega_0$, from Eq.~(\ref{eq:NG_magnon_analytical_result}) we obtain
\begin{align}\label{domega_a}
    \delta \omega_a  = &\omega_{up}(k_i) - \omega_{dn}(k_i) \approx \nonumber\\  &2 \sqrt{\frac{\beta c_s \omega_0}{2 \Delta_0 \pi^2 k_BT_c}}= 2 \sqrt{\frac{ x_c\Delta_0 \xi \omega_0 \tilde{J}_p \tilde{C}_{12}^Kc_s }{\pi}}.
\end{align}
From Eq.~(\ref{domega_a}) it follows that at low temperatures the anticrossing strength $\delta \omega_a$ is proportional to $h_0$, since $c_s \propto h_0^2$. Moreover, we can clearly see that  $\delta \omega_a$ does not depend on plasma frequency at low temperatures as well as on coherence length $\xi$, since $\tilde{J}_p \propto \xi^{-1}$. Finally, the anticrossing strength is propotional to the square root of the magnon frequency $\omega_0$.

The anticrossing strength $\delta \omega_a$ calculated numerically from Eq.~(\ref{eq:dispersion_eq_reduced}) without making use of the approximations of low temperature, low frequencies and small wavenumbers, is shown in Fig.~\ref{anti_xi} as function of all essential physical parameters. The numerical results support our analytical low-temperature findings. As it is seen from Figs.~\ref{anti_xi}(a)-(b), at low temperatures $\delta \omega_a$ does not depend on the magnon frequency $\omega_0$ and superconducting plasma frequency $\omega_p$, although this statement is violated at higher temperatures. At the same time the linear dependence of $\delta \omega_a$ on $h_0$ survives in the whole temperature range. It is the most general result, which follows directly from Eq.~(\ref{eq:dispersion_eq_reduced}) since $\delta \omega_a \propto \sqrt {M^p F_{\omega_r, \bm k}^p c_s (\bm J_p \cdot \hat x)} \propto h_0$ because the other involved quantities do not contain $h_0$.


\section{Conclusion}
\label{sec:conclusions}

We develop a linear response theory for collective excitations in a disordered 2D superconductor with full spin-momentum locking, interfaced with a thin-film ferromagnetic insulator (FI). Our approach is based on the nonequilibrium Keldysh-Usadel quasiclassical formalism, applied to the surface state of a topological superconductor (TS).

We predict that magnons in the TS and the Nambu-Goldstone phase mode in the TS can hybridize to form composite magnon – NG excitations. This coupling represents a previously unexplored form of dynamical proximity effect in S/F heterostructures. The interaction arises via interfacial exchange coupling between conduction electrons in the 2D superconducting surface state of the TS and the magnetization of the FI. A key ingredient enabling this hybridization is the spin–momentum locking inherent to the helical surface state of the TS, which enforces equal magnitudes of singlet and triplet pairing correlations. As a result, the superconducting order parameter becomes sensitive to the magnon field. Conversely, excitation of the NG mode in the TS generates an \textit{ac} supercurrent, which, due to spin–momentum locking, is accompanied by spin polarization (direct magnetoelectric effect). This current-induced spin polarization exerts a torque on the FI, thereby exciting magnons.

The magnon-NG coupling strength is investigated both analytically and numerically. We find that it scales linearly with the interface exchange constant and exhibits a square-root dependence on the magnon gap at low temperatures. The coupling shows no significant dependence on the superconducting plasma frequency in the low-temperature or high-frequency regimes. In contrast, the Higgs (amplitude) mode shows no linear coupling to the magnons and does not participate in the formation of hybrid excitations. 

The reconstructed spectra of these composite excitations offer a potential route to experimentally probe the NG mode and its coupling to magnons. Beyond its fundamental interest, the predicted effect holds promise for superconducting spintronics, as it provides a new mechanism for mutual conversion between spin signals and spinless collective superconducting excitations.

\begin{acknowledgments}
The authors are grateful to G.A. Bobkov for many useful discussions. The analytical calculations were supported by the Russian Science Foundation via the project No.~22-42-04408 and the numerical calculations were supported by Grant from the ministry of science and higher education of the Russian Federation No. 075-15-2025-010. T. K. is grateful for the support by the HSE University Basic Research Program that was used to conclude the formulation of the model. 
\end{acknowledgments}

\appendix 

\section{First-order corrections to the Green's functions}
\label{sec:appendix_A}

Here we present details of the expansion of the first-order corrections to Green's function in terms of the OP perturbation, the magnon, and the electric potential. The first order corrections to the retarded and advanced Green's functions can be expanded with respect to perturbations as follows:
\begin{align}
    [\delta \hat{g}_{\omega_r,k}^{R,A}]_{ij} =  a_{ij}^{R,A} \delta \Delta_{\omega_r, k} + b_{ij}^{R,A} \delta \Delta_{-\omega_r, -k}^* \nonumber \\
    + c_{ij}^{R,A} (\hat {\bm k} \times \delta \bm h_{\omega_r, \bm k}) \cdot\hat z + d_{ij}^{R,A} \delta\phi_{\omega_r, \bm k},
    \label{app:corrections_RA}
\end{align}
where
\begin{widetext}
\begin{eqnarray}
    a_{11}^{R,A}= - b_{22}^{R,A}  = \frac{1}{D_n^{R,A}}\left [ 2 \left(\epsilon \pm i \Gamma \right)f_0^{R,A}(\epsilon + \frac{\omega}{2}) - \Delta_{0} g^{R,A}_+ \right],\\
    a_{12}^{R,A} = b_{21}^{R,A}= \frac{1}{D_n^{R,A}}\left [ i \Delta f^{R,A}_{-} - \frac{ \hbar D}{2} k^2 f^{R,A}_{+} f^{R,A}_{-} - i g^{R,A}_+ \left( \omega + i \frac{ \hbar D}{2} k^2 g^{R,A}_-\right) \right], \\
    a_{21}^{R,A} =b_{12}^{R,A} = \frac{i \Delta f^{R,A}_-}{D_n^{R,A}} ,\\
     a_{22}^{R,A} =-b_{11}^{R,A}=\frac{1}{D_n^{R,A}}\left [ 2 \left(\epsilon \pm i \Gamma \right)f_0^{R,A}(\epsilon - \frac{\omega}{2}) - \Delta_{0} g^{R,A}_+ \right].
\end{eqnarray}
\begin{eqnarray}
     c_{11}^{R,A}= c_{22}^{R,A} =\frac{D |k_x|}{2 v_f D_n^{R,A}}\left[ \left(\frac{ \hbar D}{2} k^2 f_+^{R,A}  -  2 i  \Delta\right)  \Gamma_{x y} +
 i \left(2\left (\epsilon \pm i \Gamma \right) + i \frac{ \hbar D}{2} k^2 g_+^{R,A}\right) (\Gamma_{xx} - 1)\right], \\
      c_{12}^{R,A} = -c_{21}^{R,A} = \frac{D|k_x|}{2 v_f D_n^{R,A}}\left[  ( \omega +  i \frac{ \hbar D}{2} k^2 g_-^{R,A}) \Gamma_{x y} -  i \frac{ \hbar D}{2} k^2 f_-^{R,A} (\Gamma_{x x} - 1)\right].
\end{eqnarray}
\begin{eqnarray}
    d_{11}^{R,A} = d_{22}^{R,A} = \frac{e}{D_n^{R,A}}\left[\left( 2 (\epsilon \pm i \Gamma) + i \frac{ \hbar D}{2} k^2 g_+^{R,A} \right)g_-^{R,A}  - \left(2 \Delta + i \frac{ \hbar D}{2} k^2 f_+^{R,A}\right) f_-^{R,A} \right],\\
    d_{12}^{R,A} = - d_{21}^{R,A} = \frac{- i e \omega f_- ^{R,A} }{D_n^{R,A}},
\end{eqnarray}
and 
\begin{align}
    D_n^{R,A} &=  \left( i \Delta - \frac{ \hbar D}{2} k^2 f_0^{R,A}(\epsilon - \frac{\omega}{2})\right)^2 - \left( \frac{ \hbar D}{2} k^2 f_0^{R,A}(\epsilon + \frac{\omega}{2}) - i \Delta \right)^2 \nonumber \\ 
    &- \left(2 \left (\epsilon \pm i \Gamma \right) + i \frac{ \hbar D}{2} k^2 g_+^{R,A}\right) \left( \omega + i \frac{ \hbar D}{2} k^2 g_-^{R,A}\right),
\end{align}
\begin{eqnarray}
    \Gamma_{xx} = f_0^{R,A} (\epsilon + \frac{\omega}{2}) f_0^{R,A} (\epsilon - \frac{\omega}{2}) + g_0^{R,A} (\epsilon + \frac{\omega}{2}) g_0^{R,A} (\epsilon - \frac{\omega}{2}),
\end{eqnarray}    
\begin{eqnarray}
\Gamma_{xy} = f_0^{R,A} (\epsilon + \frac{\omega}{2}) g_0^{R,A} (\epsilon - \frac{\omega}{2}) + g_0^{R,A} (\epsilon + \frac{\omega}{2}) f_0^{R,A} (\epsilon - \frac{\omega}{2}),
\end{eqnarray} 
\begin{eqnarray}
    f[g]_\pm = f[g]_0^{R,A} (\epsilon + \frac{\omega}{2}) \pm f[g]_0^{R,A} (\epsilon - \frac{\omega}{2}).
\end{eqnarray} 

The first order corrections to the Keldysh Green's functions can be also expanded with respect to perturbations. They can be compactly written as follows:
\begin{align}
&\left[\delta g_{\omega_r,k}^{K}\right]_{11} = \frac{1}{D_n^K} \left(-a_2 b_2 + a_3 b_3 + a_4 b_1\right), \label{g_K_11} \\
&\left[\delta f_{\omega_r,k}^{K}\right]_{12} = \frac{1}{D_n^K} \left(a_1 b_2 - a_2 b_1 + a_3 b_4\right), \label{g_K_12} \\
&\left[\delta f_{\omega_r,k}^{K}\right]_{21} = \frac{1}{D_n^K} \left(-a_1 b_3 - a_2 b_4 + a_3 b_1\right), \label{g_K_21} \\
&\left[\delta g_{\omega_r,k}^{K}\right]_{22} = \frac{1}{D_n^K} \left(-a_2 b_3 + a_3 b_2 - a_4 b_4\right), 
\label{g_K_22}
\end{align}
where $a_1-a_4$ are of zero order with respect to the perturbations and take the form:
\begin{align}
&a_1 = \omega - 2 i \Gamma - i { \hbar D} k^2 g_0^R (\epsilon - \frac{\omega}{2}),\quad
a_2 = i \Delta , \nonumber \\
&a_3 = -i \Delta + { \hbar D} k^2 f_0^R (\epsilon - \frac{\omega}{2}), \quad
a_4 = -\left( 2 \epsilon + i { \hbar D} k^2 g_0^R (\epsilon - \frac{\omega}{2})\right),
\end{align}
$D_n^K = a_1 a_4 - a_2^2 + a_3^2$, and $b_1-b_4$ contain first order terms with respect to perturbations:
\begin{align}
    b_1 = &- \delta \Delta_{\omega_r,k}f_0^K (\epsilon + \frac{\omega}{2}) + \delta \Delta^*_{-\omega_r,-k}f_0^K (\epsilon - \frac{\omega}{2}) - i \frac{\delta h}{2 v_f}  D k_x \Gamma_{xx}^K - g_-^K \delta \phi_{\omega_r,k} \nonumber\\ 
    &- 2 i \Gamma \left( \tanh{\frac{\epsilon + \omega/2}{2 T}} \delta {g}^R_{11}  -\tanh{\frac{\epsilon - \omega/2}{2 T}} \delta {g}^A_{11} \right)   - { \hbar D} k^2 \left( f_0^K (\epsilon - \frac{\omega}{2}) \delta f ^A_{21} - i g_0^K (\epsilon - \frac{\omega}{2}) \delta g^A_{11}\right),
\end{align}
\begin{align}
    b_2 = &- i \delta \Delta_{\omega_r,k} g_+^K + \frac{\delta h}{2 v_f} Dk_x \Gamma_{xy}^K + 2 i \Gamma \left( \tanh{\frac{\epsilon + \omega/2}{2 T}} \delta {f}^R_{12}  +\tanh{\frac{\epsilon - \omega/2}{2 T}} \delta {f}^A_{12} \right) - i f_-^K \delta \phi_{\omega_r,k}   \nonumber \\
    &- { \hbar D} k^2 \left( f_0^K (\epsilon - \frac{\omega}{2}) \delta g ^A_{22}  - i g_0^K (\epsilon - \frac{\omega}{2}) \delta f^A_{12}\right),
\end{align}
\begin{align}
    b_3 = &i \delta \Delta^*_{-\omega_r,-k} g_+^K + \frac{\delta h}{2 v_f}D k_x \Gamma_{xy}^K - 2 i \Gamma \left( \tanh{\frac{\epsilon + \omega/2}{2 T}} \delta {f}^R_{21}  + \tanh{\frac{\epsilon - \omega/2}{2 T}} \delta {f}^A_{21} \right)  - i f_-^K \delta \phi_{\omega_r,k} \nonumber\\ 
    &- { \hbar D} k^2 \left(f_0^K (\epsilon - \frac{\omega}{2}) \delta g ^A_{11} + i g_0^K (\epsilon - \frac{\omega}{2}) \delta f^A_{21}\right),
\end{align}
\begin{align}
    b_4 = &\delta \Delta_{\omega_r,k} f_0^K (\epsilon - \frac{\omega}{2}) - \delta \Delta^*_{-\omega_r,-k}f_0^K (\epsilon + \frac{\omega}{2}) + i \frac{\delta h}{2 v_f}D k_x \Gamma_{xx}^K +  g_-^K \delta \phi_{\omega_r,k}\nonumber\\
    &+ 2 i \Gamma \left( \tanh{\frac{\epsilon + \omega/2}{2 T}} \delta {g}^R_{22}  -\tanh{\frac{\epsilon - \omega/2}{2 T}} \delta {g}^A_{22} \right)    -  { \hbar D} k^2 \left( f_0^K (\epsilon - \frac{\omega}{2}) \delta f ^A_{12} + i g_0^K (\epsilon - \frac{\omega}{2}) \delta g^A_{22}\right),
\end{align}
\begin{eqnarray}
\Gamma_{xy}^K =  g_0^K(\epsilon - \frac{\omega}{2}) f_0 ^A (\epsilon + \frac{\omega}{2}) + g_0^R (\epsilon - \frac{\omega}{2}) f_0^K (\epsilon + \frac{\omega}{2})  
+ f_0^K (\epsilon - \frac{\omega}{2}) g_0^A (\epsilon + \frac{\omega}{2})  + f_0^R (\epsilon - \frac{\omega}{2})g_0^K (\epsilon + \frac{\omega}{2}) ,
\label{eq:Gamma_xy_K}
\end{eqnarray}
\begin{eqnarray}
 \Gamma_{xx}^K = f_0^K(\epsilon - \frac{\omega}{2}) f_0^A (\epsilon + \frac{\omega}{2}) + g_0^R (\epsilon - \frac{\omega}{2}) g_0^K (\epsilon + \frac{\omega}{2})  + 
 g_0^K (\epsilon - \frac{\omega}{2}) g_0^A (\epsilon + \frac{\omega}{2})  + f_0^R (\epsilon - \frac{\omega}{2}) f_0^K (\epsilon + \frac{\omega}{2}) .
\label{eq:Gamma_xx_K}
\end{eqnarray}
Eqs.~(\ref{g_K_11})-(\ref{g_K_22}) can be rewritten in the form of Eq.~(\ref{corrections_K}), however the explicit expressions for the coefficients $a_{ij}^{K}$, $b_{ij}^{K}$, $c_{ij}^{K}$ and $d_{ij}^{K}$ are rather cumbersome and the representation used here seems more convenient.

\section{Amplitude response to the magnon}
\label{app:amplitude_response}
\begin{eqnarray}
    F_{\omega_r,\bm k}^a =\frac{-C^{K}_{12} (\omega, \bm k) \left(A_{12}^{K*} (-\omega^*, -\bm k) - 1 - B_{12}^{K*} (-\omega^*, -\bm k)\right)+ C^{K*}_{12} (-\omega^*, -\bm k) \left(A_{12}^{K} (\omega, \bm k) - 1 - B_{12}^{K} (\omega, \bm k)\right)}{i \cdot det\left[\hat{M}_{\Delta\Delta}\right]} .
\end{eqnarray}
In order to prove the absence of the amplitude response to the magnon, we will find symmetry relations between the retarded and advanced components upon the operation of complex conjugation and substitution $\omega, \bm k \rightarrow -\omega^*, - \bm k$.

First, we perform the complex conjugation and substitute $\omega, \bm k \rightarrow -\omega^*, - \bm k$ in Eq.\eqref{corrections_RA} and assume $\Gamma \rightarrow 0$, which takes the form
\begin{align}
    [\delta \hat{g}_{-\omega_r,-\bm k}^{R*}]_{ij} =  a_{ij}^{R*}(-\omega^*, -\bm k) \delta \Delta_{-\omega_r, -\bm k}^* + b_{ij}^{R*}(-\omega^*, -\bm k) \delta \Delta_{\omega_r,k} 
    - c_{ij}^{R*}(-\omega^*, -\bm k) (\hat {\bm k} \times \delta \bm h_{-\omega_r, -\bm k}) \cdot\hat z,
    \label{corrections_RA_star}
\end{align}
where
\begin{align}
    a_{11}^{R*}(-\omega^*, -\bm k) &= - b_{22}^{R*}(-\omega^*, -\bm k)  = -\frac{1}{D_n^{A}} \left [- 2 \left(\epsilon - i \Gamma \right)f_0^{A}(\epsilon - \frac{\omega}{2}) + \Delta_{0} g^{A}_+ \right] = a_{22}^{A},\\
    a_{12}^{R*}(-\omega^*, -\bm k) &= b_{21}^{R*}(-\omega^*, -\bm k)= -\frac{1}{D_n^{A}}\left [ -i \Delta f^{A}_{-} + \frac{\hbar D}{2} k^2 f^{A}_{+} f^{A}_{-} - i g^{A}_+ \left( -\omega - i \frac{\hbar D}{2} k^2 g^{A}_-\right) \right] = a_{12}^{A}, \\
    a_{21}^{R*}(-\omega^*, -\bm k) &=b_{12}^{R*}(-\omega^*, -\bm k) = \frac{i \Delta f^{A}_-}{D_n^{A}} =a_{21}^{A},\\
     a_{22}^{R*} (-\omega^*, -\bm k)&=-b_{11}^{R*}(-\omega^*, -\bm k)=-\frac{1}{D_n^{A}}\left [ -2 \left(\epsilon - i \Gamma \right)f_0^{A}(\epsilon + \frac{\omega}{2}) + \Delta_{0} g^{A}_+ \right] = a_{11}^{A}.
    \end{align}
\begin{align}
     c_{11}^{R*}(-\omega^*, -\bm k) &= c_{22}^{R*}(-\omega^*, -\bm k) =-\frac{ D|k_x|}{2 v_f D_n^{A}}\left[ \left(-\frac{\hbar D}{2} k^2 f_+^{A}  +  2 i  \Delta\right)  \Gamma_{x y}^A -
 i \left(2\left (\epsilon - i \Gamma \right) + i \frac{\hbar D}{2} k^2 g_+^{A}\right) (\Gamma_{xx} - 1)\right] = c_{11}^A, \\
      c_{12}^{R*}(-\omega^*, -\bm k) &= -c_{21}^{R *}(-\omega^*, -\bm k) = -\frac{D|k_x|}{2 v_f D_n^{A}} \left[( -\omega -  i \frac{\hbar D}{2} k^2 g_-^{A}) \Gamma_{x y} +  i \frac{\hbar D}{2} k^2 f_-^{A} (\Gamma_{x x} - 1)\right] = c_{12}^A,
\end{align}
where we used
\begin{eqnarray}
\left(f[g]_0^R\left(\epsilon \mp \frac{\omega^*}{2}\right)\right)^* = - f[g]_0^A(\epsilon \mp \frac{\omega}{2}), \quad \left(\Gamma_{xx}^{R} ( - \omega^*)\right)^* = \Gamma_{xx}^{A}, \quad \left(\Gamma_{xy}^{R} (-\omega^*)\right)^* = \Gamma_{xy}^{A}, \quad \left(D_n^R (-\omega^*, -\bm k)\right)^* = -D_n^A. \nonumber
\end{eqnarray}
Let us rewrite the coefficients as
\begin{align}
    a_{11}^{R*}(-\omega^*, -\bm k) &= a_{22}^{A}(\omega, \bm k),\\
    a_{12}^{R*}(-\omega^*, -\bm k) &= a_{12}^{A}(\omega, \bm k), \\
    a_{21}^{R*}(-\omega^*, -\bm k) &= a_{21}^{A}(\omega, \bm k), \\
     a_{22}^{R*} (-\omega^*, -\bm k)&= a_{11}^{A}(\omega, \bm k), \\
     c_{11}^{R*}(-\omega^*, -\bm k)& = c_{11}^A(\omega, \bm k), \\
     c_{12}^{R*}(-\omega^*, -\bm k) &=c_{12}^A(\omega, \bm k).
\end{align}
As we can notice from above, the complex conjugation and $(\omega, \bm k)$ inversion leads to the symmetry relationships between retarded and advanced coefficients. However, as one can recognize, the Usadel equation for the Keldysh components written in a standard way is asymmetrical in a sense that it contains only $\delta\hat{g}^K$, $\delta\hat{g}^A$, but not $\delta\hat{g}^R$ (see Eq. \eqref{keldysh_corr_solution}). Therefore, to simplify the analysis it is useful to utilize the normalization condition for the Keldysh Green's function
\begin{eqnarray}\label{norm_Keld}
    \hat {g}_0^R(\epsilon - \frac{\omega}{2}) \delta \hat g^K+ \hat {g}_0^K(\epsilon - \frac{\omega}{2}) \delta \hat g^A + \delta \hat g^R  \hat {g}_0^K(\epsilon + \frac{\omega}{2})  + \delta \hat g^K  \hat {g}_0^A(\epsilon + \frac{\omega}{2}) =0.
\end{eqnarray}
First, we consider the Keldysh equation written in a standard way as in Eq.\eqref{keldysh_corr_solution}, then we derive the second (identical) equation using normalization condition \eqref{norm_Keld}, substituting
\begin{eqnarray}
    \hat {g}_0^R(\epsilon - \frac{\omega}{2}) \delta \hat g^K+ \hat {g}_0^K(\epsilon - \frac{\omega}{2}) \delta \hat g^A = - \left( \delta \hat g^R  \hat {g}_0^K(\epsilon + \frac{\omega}{2})  + \delta \hat g^K  \hat {g}_0^A(\epsilon + \frac{\omega}{2}) \right)
\end{eqnarray}
in the LHS of Eq. \eqref{keldysh_corr_solution} and take complex conjugation together with $(\omega, \bm k)$ inversion.

From the first equation, Keldysh off-diagonal component $\delta{g}^K_{12}$ can be written in the following form
\begin{eqnarray}
    [\delta \hat{g}_{\omega_r,\bm k}^{K}]_{12} =  a_{12}^{K}(\omega,\bm k) \delta \Delta_{\omega_r, \bm k} + b_{12}^{K}(\omega, \bm k) \delta \Delta_{-\omega_r,-\bm k}^* 
    + c_{12}^{K}(\omega,\bm k) (\hat {\bm k} \times \delta \bm h_{\omega_r, \bm k}) \cdot\hat z,
\end{eqnarray}

where
\begin{align}
a_{12}^K &= \frac{1}{D_n^K} \left[i \Delta f_-^K + {\hbar D} k^2 f_0^R(\epsilon - \frac{\omega}{2})f_0^K(\epsilon - \frac{\omega}{2}) - i g_+^K \left( \omega - i {\hbar D} k^2 g_0^R(\epsilon - \frac{\omega}{2})\right) 
 + c_{\delta\Delta}^A(\omega, \bm k)\right], \nonumber \\
 b_{12}^K &= \frac{1}{D_n^K} \left[ i \Delta f_-^K - \hbar {D} k^2 f_0^R(\epsilon - \frac{\omega}{2})f_0^K(\epsilon + \frac{\omega}{2}) 
  + c_{\delta\Delta^*}^A(\omega, \bm k)\right], \nonumber \\
 c_{12}^K &= \frac{D|k_x|}{2 v_f D_n^K} \left[  i {\hbar D} k^2 f_0^R(\epsilon - \frac{\omega}{2}) \Gamma_{xx}^K + \left( \omega -i {\hbar D} k^2g_0^R(\epsilon - \frac{\omega}{2})\right) \Gamma_{xy}^K+ c_{\delta h}^A(\omega, \bm k)\right],
\end{align}
Here, coefficients $c$ are defined as follows
\begin{align}
    c_{\delta \Delta}^A(\omega, \bm k) &= \Pi_{11}^A a_{11}^A + \Pi_{12}^A a_{12}^A + \Pi_{21}^A a_{21}^A + \Pi_{22}^A a_{22}^A, \\
    c_{\delta \Delta^*}^A(\omega, \bm k) &= -\Pi_{11}^A a_{22}^A + \Pi_{12}^A a_{21}^A + \Pi_{21}^A a_{12}^A - \Pi_{22}^A a_{11}^A,\\
    c_{\delta h}^A(\omega, \bm k) &= \Pi_{11}^A c_{11}^A + \Pi_{12}^A c_{12}^A  -\Pi_{21}^A c_{12}^A + \Pi_{22}^A c_{11}^A.
\end{align}

with compact notations
\begin{align}
   \Pi_{11}^A(\omega, \bm k) &=  {\hbar D} k^2 g_0^K(\epsilon - \frac{\omega}{2}) \Delta, \\
    \Pi_{12}^A(\omega, \bm k) &= i {\hbar D} k^2 \left[g_0^K(\epsilon - \frac{\omega}{2})\left( \omega - i {\hbar D} k^2 g_0^R(\epsilon - \frac{\omega}{2})\right) - i f_0^K(\epsilon - \frac{\omega}{2}) \left( i \Delta -{\hbar D}k^2f_0^R(\epsilon - \frac{\omega}{2}) \right)\right],\\
    \Pi_{21}^A(\omega, \bm k) &= i {\hbar D} k^2 f_0^K(\epsilon - \frac{\omega}{2}) \Delta ,\\
    \Pi_{22}^A(\omega, \bm k) &=-{\hbar D} k^2 \left[f_0^K(\epsilon - \frac{\omega}{2})\left( \omega -i {\hbar D} k^2 g_0^R(\epsilon - \frac{\omega}{2})\right) - i g_0^K(\epsilon - \frac{\omega}{2}) \left( i \Delta - {\hbar D} k^2f_0^R(\epsilon - \frac{\omega}{2}) \right)\right].
\end{align}
From the second Keldysh equation we can obtain
\begin{eqnarray}
    [\delta \hat{g}_{-\omega_r,-\bm k}^{K}]_{12}^* =  a_{12}^{K*}(-\omega^*,-\bm k) \delta \Delta_{-\omega_r, -\bm k}^* + b_{12}^{K*}(-\omega^*, -\bm k) \delta \Delta_{\omega_r,\bm k} 
    - c_{12}^{K*}(-\omega^*,-\bm k) (\hat {\bm k} \times \delta \bm h_{-\omega_r, -\bm k}) \cdot\hat z,
\end{eqnarray}
\begin{align}
a_{12}^{K*}(-\omega^*,-\bm k) &= -\frac{1}{D_n^K} \left[i \Delta f_-^K + \hbar {D} k^2 f_0^R(\epsilon - \frac{\omega}{2})f_0^K(\epsilon - \frac{\omega}{2}) - i g_+^K \left( \omega -i {\hbar D} k^2 g_0^R(\epsilon - \frac{\omega}{2})\right) 
 + c_{\delta\Delta}^{R*}(-\omega^*, -\bm k)\right], \nonumber \\
 b_{12}^{K*}(-\omega^*,-\bm k) &= -\frac{1}{D_n^K} \left[ i \Delta f_-^K - {\hbar D} k^2 f_0^R(\epsilon - \frac{\omega}{2})f_0^K(\epsilon + \frac{\omega}{2}) 
  + c_{\delta\Delta^*}^{R*}(-\omega^*, -\bm k)\right], \nonumber \\
 c_{12}^{K*}(-\omega^*,-\bm k) &= -\frac{D |k_x|}{2v_fD_n^K} \left[ - i {\hbar D} k^2 f_0^R(\epsilon - \frac{\omega}{2}) \Gamma_{xx}^K - \left( \omega - i {\hbar D} k^2g_0^R(\epsilon - \frac{\omega}{2})\right) \Gamma_{xy}^K+ c_{\delta h}^{R*}(-\omega^*, -\bm k)\right],
\end{align}
with coefficients
\begin{align}
    c_{\delta \Delta}^{R*}(-\omega^*, -\bm k) &= \Pi_{22}^A a_{11}^A + \Pi_{12}^A a_{12}^A + \Pi_{21}^A a_{21}^A + \Pi_{11}^A a_{22}^A, \\
    c_{\delta \Delta^*}^{R*}(-\omega^*, -\bm k) &= -\Pi_{22}^A a_{22}^A + \Pi_{12}^A a_{21}^A + \Pi_{21}^A a_{12}^A - \Pi_{11}^A a_{11}^A,\\
    c_{\delta h}^{R*}(-\omega^*, -\bm k) &= \Pi_{22}^A c_{11}^A + \Pi_{12}^A c_{12}^A  -\Pi_{21}^A c_{12}^A + \Pi_{11}^A c_{11}^A,
\end{align}

where we used the following expressions
\begin{align}\label{RA_Pi}
    \Pi_{11}^{R*}(-\omega^*, -\bm k) &= \Pi_{22}^A(\omega, \bm k) , \\
    \Pi_{12}^{R*}(-\omega^*, -\bm k) &=\Pi_{12}^A(\omega, \bm k) ,\\
    \Pi_{21}^{R*}(-\omega^*, -\bm k) &=\Pi_{21}^A(\omega, \bm k) ,\\
    \Pi_{22}^{R*}(-\omega^*, -\bm k) &=\Pi_{11}^A(\omega, \bm k) , \\
    \left(D_n^K (-\omega^*, - \bm k)\right)^* &= -D_n^K.
\end{align}

The Coulomb interaction leads to the renormalization of $a^{K}$, $b^{K}$ and $c^{K}$. Thus, one needs to determine the symmetry relations for the renormalized coefficients in Eq.\eqref{dg_K_poisson}. In this case, we must find the transformations for diagonal components of the Keldysh Green's functions, including

\begin{align}
    a_{11}^K = \frac{1}{ D_n^K} \left[2 f_0^K(\epsilon + \frac{\omega}{2})  \left(\epsilon + i \frac{\hbar D}{2} k^2 g_0^R(\epsilon - \frac{\omega}{2})\right) - g_+^K \Delta + c_{\delta\Delta}^{11}(\omega, \bm k)\right],\\
    b_{11}^K = \frac{1}{ D_n^K} \left[ -2 f_0^K(\epsilon - \frac{\omega}{2})\left(\epsilon + i \frac{\hbar D}{2} k^2 g_0^R(\epsilon - \frac{\omega}{2})\right)- i g_+^K\left( i \Delta - {\hbar D} k^2f_0^R(\epsilon - \frac{\omega}{2}) \right)+ c_{\delta\Delta^*}^{11}(\omega, \bm k)\right],\\
    c_{11}^K = \frac{D|k_x|}{ 2 v_fD_n^K} \left[ 2i\left( \epsilon +i \frac{\hbar D}{2} k^2g_0^R(\epsilon - \frac{\omega}{2})\right) \Gamma_{xx}^K -  2  \left( i \Delta - \frac{\hbar D}{2} k^2f_0^R(\epsilon - \frac{\omega}{2}) \right) \Gamma_{xy}^K  + c_{\delta h}^{11}(\omega, \bm k)\right],\\
    d_{11}^K = \frac{e}{ D_n^K} \left[2\left( \epsilon +i \frac{\hbar D}{2} k^2g_0^R(\epsilon - \frac{\omega}{2})\right) g_-^K - 2\left( \Delta + i \frac{\hbar D}{2} k^2f_0^R(\epsilon - \frac{\omega}{2}) \right)f^K_- 
+c_{\delta \phi}^{11}(\omega, \bm k) \right],\\
\end{align}
and
\begin{align}
    a_{22}^K = \frac{1}{ D_n^K} \left[2 f_0^K(\epsilon - \frac{\omega}{2})\left(\epsilon + i \frac{\hbar D}{2} k^2 g_0^R(\epsilon - \frac{\omega}{2})\right)+ i g_+^K\left( i \Delta -{\hbar D} k^2f_0^R(\epsilon - \frac{\omega}{2}) \right) + c_{\delta\Delta}^{22}(\omega, \bm k)\right],\\
    b_{22}^K = \frac{1}{ D_n^K} \left[ -2 f_0^K(\epsilon + \frac{\omega}{2})  \left(\epsilon + i \frac{\hbar D}{2} k^2 g_0^R(\epsilon - \frac{\omega}{2})\right) + g_+^K \Delta+ c_{\delta\Delta^*}^{22}(\omega, \bm k)\right],\\
    c_{22}^K = \frac{D|k_x|}{ 2 v_fD_n^K} \left[ 2i\left( \epsilon +i \frac{\hbar D}{2} k^2g_0^R(\epsilon - \frac{\omega}{2})\right) \Gamma_{xx}^K -  2  \left( i \Delta - \frac{\hbar D}{2} k^2f_0^R(\epsilon - \frac{\omega}{2}) \right) \Gamma_{xy}^K  + c_{\delta h}^{22}(\omega, \bm k)\right],\\
    d_{22}^K = \frac{e}{ D_n^K} \left[2\left( \epsilon +i \frac{\hbar D}{2} k^2g_0^R(\epsilon - \frac{\omega}{2})\right) g_-^K - 2\left( \Delta + i \frac{\hbar D}{2} k^2f_0^R(\epsilon - \frac{\omega}{2}) \right)f^K_- 
+c_{\delta \phi}^{22}(\omega, \bm k) \right],\\
\end{align}
where
\begin{align}
     c_{\delta\Delta}^{11(22)}(\omega, \bm k) = \Pi_{11}^{11(22),A} a_{11}^A + \Pi_{12}^{11(22),A} a_{12}^A + \Pi_{21}^{11,A} a_{21}^A + \Pi_{22}^{11(22),A} a_{22}^A, \\
    c_{\delta\Delta^*}^{11(22)}(\omega, \bm k) = -\Pi_{11}^{11(22),A} a_{22}^A + \Pi_{12}^{11(22),A} a_{21}^A + \Pi_{21}^{11(22),A} a_{12}^A - \Pi_{22}^{11(22),A} a_{11}^A, \\
     c_{\delta h}^{11(22)}(\omega, \bm k) = \Pi_{11}^{11(22),A} c_{11}^A + \Pi_{12}^{11(22),A} c_{12}^A - \Pi_{21}^{11(22),A} c_{12}^A +\Pi_{22}^{11(22),A} c_{22}^A, \\
    c_{\delta \phi}^{11(22)}(\omega, \bm k) = \Pi_{11}^{11(22),A} d_{11}^A + \Pi_{12}^{11(22),A} d_{12}^A + \Pi_{21}^{11(22),A} d_{21}^A + \Pi_{22}^{11(22),A} d_{22}^A,
\end{align}

\begin{align}
    \Pi_{11}^{11,A}(\omega, \bm k) &= \Pi_{22}^{22,A}(\omega, \bm k)=  {\hbar D} k^2\left[f_0^K(\epsilon - \frac{\omega}{2}) \left( i \Delta -{\hbar D} k^2f_0^R(\epsilon - \frac{\omega}{2}) \right)  - 2 i g_0^K(\epsilon - \frac{\omega}{2}) \left( \epsilon +  i \frac{\hbar D}{2} k^2g_0^R(\epsilon - \frac{\omega}{2}) \right)\right] , \\
    \Pi_{12}^{11,A}(\omega, \bm k) &= -\Pi_{21}^{22,A}(\omega, \bm k)= {\hbar D} k^2 g_0^K(\epsilon - \frac{\omega}{2}) \Delta   , \\
    \Pi_{21}^{11,A}(\omega, \bm k) &= -\Pi_{12}^{22,A}(\omega, \bm k)=  i {\hbar D} k^2\left[g_0^K(\epsilon - \frac{\omega}{2}) \left( i \Delta - {\hbar D} k^2f_0^R(\epsilon - \frac{\omega}{2}) \right)  - 2 i  f_0^K(\epsilon - \frac{\omega}{2}) \left( \epsilon + i \frac{\hbar D}{2} k^2g_0^R(\epsilon - \frac{\omega}{2}) \right)\right],\\
    \Pi_{22}^{11,A}(\omega, \bm k) &= \Pi_{11}^{22,A}(\omega, \bm k)=i {\hbar D} k^2 f_0^K(\epsilon - \frac{\omega}{2}) \Delta .
\end{align}
Using the same trick, involving the normalization condition \eqref{norm_Keld}, we arrive at the corresponding expressions
\begin{align}
a_{11}^{K*} ( -\omega^*, - \bm k)= -\frac{1}{ D_n^K} \left[ 2 f_0^K(\epsilon - \frac{\omega}{2})\left(\epsilon + i \frac{\hbar D}{2} k^2 g_0^R(\epsilon - \frac{\omega}{2})\right)+ i g_+^K\left( i \Delta - {\hbar D} k^2f_0^R(\epsilon - \frac{\omega}{2}) \right)+ c_{\delta\Delta}^{11*}(-\omega^*, -\bm k)\right],\\
b_{11}^{K*} ( -\omega^*, - \bm k)= -\frac{1}{ D_n^K} \left[-2 f_0^K(\epsilon + \frac{\omega}{2})  \left(\epsilon + i \frac{\hbar D}{2} k^2 g_0^R(\epsilon - \frac{\omega}{2})\right) + g_+^K \Delta + c_{\delta\Delta^*}^{11*}(-\omega^*, -\bm k)\right], \\
c_{11}^{K*} ( -\omega^*, - \bm k) = -\frac{D|k_x|}{ 2 v_fD_n^K} \left[ 2i\left( \epsilon +i \frac{\hbar D}{2} k^2g_0^R(\epsilon - \frac{\omega}{2})\right) \Gamma_{xx}^K -  2  \left( i \Delta - \frac{\hbar D}{2} k^2f_0^R(\epsilon - \frac{\omega}{2}) \right) \Gamma_{xy}^K  + c_{\delta h}^{11*}(-\omega^*, -\bm k)\right],\\
d_{11}^{K*} ( -\omega^*, - \bm k)= -\frac{e}{ D_n^K} \left[-2\left( \epsilon +i \frac{\hbar D}{2} k^2g_0^R(\epsilon - \frac{\omega}{2})\right) g_-^K + 2\left( \Delta + i \frac{\hbar D}{2} k^2f_0^R(\epsilon - \frac{\omega}{2}) \right)f^K_- 
+c_{\delta \phi}^{11*}(-\omega^*, -\bm k) \right],
\end{align}
and
\begin{align}
    a_{22}^{K*} (-\omega^*, - \bm k) = -\frac{1}{ D_n^K} \left[ 2 f_0^K(\epsilon + \frac{\omega}{2})  \left(\epsilon + i \frac{\hbar D}{2} k^2 g_0^R(\epsilon - \frac{\omega}{2})\right) - g_+^K \Delta+ c_{\delta\Delta}^{22*}(-\omega^*, -\bm k)\right],\\
    b_{22}^{K*} ( -\omega^*, - \bm k) = -\frac{1}{ D_n^K} \left[-2 f_0^K(\epsilon - \frac{\omega}{2})\left(\epsilon + i \frac{\hbar D}{2} k^2 g_0^R(\epsilon - \frac{\omega}{2})\right)- i g_+^K\left( i \Delta - {\hbar D} k^2f_0^R(\epsilon - \frac{\omega}{2}) \right) + c_{\delta\Delta^*}^{22*}(-\omega^*, -\bm k)\right],\\
    c_{22}^{K*} ( -\omega^*, - \bm k) = -\frac{D|k_x|}{ 2 v_fD_n^K} \left[ 2i\left( \epsilon +i \frac{\hbar D}{2} k^2g_0^R(\epsilon - \frac{\omega}{2})\right) \Gamma_{xx}^K -  2  \left( i \Delta - \frac{\hbar D}{2} k^2f_0^R(\epsilon - \frac{\omega}{2}) \right) \Gamma_{xy}^K  + c_{\delta h}^{22*}(-\omega^*, -\bm k)\right],\\
    d_{22}^{K*} ( -\omega^*, - \bm k) = -\frac{e}{ D_n^K} \left[-2\left( \epsilon +i \frac{\hbar D}{2} k^2g_0^R(\epsilon - \frac{\omega}{2})\right) g_-^K +2\left( \Delta + i \frac{\hbar D}{2} k^2f_0^R(\epsilon - \frac{\omega}{2}) \right)f^K_- 
+c_{\delta \phi}^{22*}(-\omega^*, -\bm k) \right],
\end{align}

\begin{align}
     c_{\delta\Delta}^{11(22)*}(-\omega^*, -\bm k) = \Pi_{11}^{11(22),A} a_{22}^A + \Pi_{12}^{22(11),A} a_{12}^A + \Pi_{21}^{22(11),A} a_{21}^A + \Pi_{22}^{11(22),A} a_{11}^A,\\
    c_{\delta\Delta^*}^{11(22)*}(-\omega^*, -\bm k) = -\Pi_{11}^{11(22),A} a_{22}^A + \Pi_{12}^{22(11),A} a_{21}^A + \Pi_{21}^{22(11),A} a_{12}^A - \Pi_{22}^{11(22),A} a_{11}^A, \\
     c_{\delta h}^{11(22)*}(-\omega^*,- \bm k) = \Pi_{11}^{11(22),A} c_{11}^A +\Pi_{12}^{22(11),A} c_{12}^A - \Pi_{21}^{22(11),A} c_{12}^A +\Pi_{22}^{11(22),A} c_{22}^A, \\
    c_{\delta \phi}^{11(22)*}(-\omega^*,- \bm k) =- \Pi_{11}^{11(22),A} d_{11}^A -\Pi_{12}^{22(11),A} d_{12}^A - \Pi_{21}^{22(11),A} d_{21}^A - \Pi_{22}^{11(22),A} d_{22}^A,
\end{align}
where we have utilized
\begin{align}
    \Pi_{11}^{11,R*}(-\omega^*, -\bm k) &=\Pi_{22}^{22,R*}(-\omega^*, -\bm k) =\Pi_{11}^{11,A}(\omega, \bm k) =\Pi_{22}^{22,A}(\omega, \bm k), \\
    \Pi_{21}^{11,R*}(-\omega^*, -\bm k)&=-\Pi_{12}^{22,R*}(-\omega^*, -\bm k)=-\Pi_{12}^{11,A}(\omega, \bm k) = \Pi_{21}^{22,A}(\omega, \bm k), \\
    \Pi_{12}^{11,R*}(-\omega^*, -\bm k) &=-\Pi_{21}^{22,R*}(-\omega^*, -\bm k) = -\Pi_{21}^{11,A}(\omega, \bm k) = \Pi_{12}^{22,A}(\omega, \bm k),\\
    \Pi_{22}^{11,R*}(-\omega^*, -\bm k) &=\Pi_{11}^{22,R*}(-\omega^*, -\bm k) =\Pi_{22}^{11,A}(\omega, \bm k) = \Pi_{11}^{22,A}(\omega, \bm k).
\end{align}
and 
\begin{align}
    d_{11}^{R*}(-\omega^*, -\bm k) = -d_{11}^{A}(\omega, \bm k), \\
    d_{12}^{R*}(-\omega^*, -\bm k) = -d_{12}^{A}(\omega, \bm k), \\
    d_{21}^{R*}(-\omega^*, -\bm k) = -d_{21}^{A}(\omega, \bm k), \\
    d_{22}^{R*}(-\omega^*, -\bm k) = -d_{22}^{A}(\omega, \bm k).
\end{align}
The diagonal coefficients transform in the following way
\begin{align}
    a_{11}^K(\omega, \bm k) + a_{22}^K(\omega, \bm k) &=  -\left[a_{11}^{K*}(-\omega^*, -\bm k) + a_{22}^{K*}(-\omega^*, -\bm k)\right],\\
    b_{11}^K(\omega, \bm k) + b_{22}^K(\omega, \bm k) &=  -\left[b_{11}^{K*}(-\omega^*, -\bm k) + b_{22}^{K*}(-\omega^*, -\bm k)\right],\\
    c_{11}^K(\omega, \bm k) + c_{22}^K(\omega, \bm k) &=  -\left[c_{11}^{K*}(-\omega^*, -\bm k) + c_{22}^{K*}(-\omega^*, -\bm k)\right],\\
    d_{11}^K(\omega, \bm k) + d_{22}^K(\omega, \bm k) &= d_{11}^{K*}(-\omega^*, -\bm k) + d_{22}^{K*}(-\omega^*, -\bm k),
\end{align}
which means that
\begin{eqnarray}
    \label{n_sym_rel}
    n_{\delta\Delta}(\omega, \bm k) &= -n^*_{\delta\Delta}(-\omega^*, -\bm k),\\ n_{\delta\Delta^*}(\omega, \bm k) &= -n^*_{\delta\Delta^*}(-\omega^*, -\bm k), \\ n_{\delta h}(\omega, \bm k) &= -n^*_{\delta h}(-\omega^*, -\bm k), \\ n_{\delta \phi}(\omega, \bm k) &= n^*_{\delta\phi}(-\omega^*, -\bm k).
\end{eqnarray}
Finally, we obtain the expressions for $d_{12}^K(\omega, \bm k) $ and $d_{12}^{K*}(-\omega^*,-\bm k)$
\begin{align}
     d_{12}^K(\omega, \bm k) &= \frac{e}{D_n^K} \left[ {\hbar D} k^2 f_0^R(\epsilon - \frac{\omega}{2}) g_-^K - i \left(\omega -  i {\hbar D} k^2 g_0^R(\epsilon - \frac{\omega}{2})\right) f_-^K + c_{\delta \phi}^{12} (\omega, \bm k)\right], \\
     d_{12}^{K*}(-\omega^*, -\bm k) &= -\frac{e}{D_n^K} \left[ -{\hbar D} k^2 f_0^R(\epsilon - \frac{\omega}{2}) g_-^K + i \left(\omega - i {\hbar D} k^2 g_0^R(\epsilon - \frac{\omega}{2})\right) f_-^K + c_{\delta \phi}^{12*} (-\omega^*, -\bm k)\right].
\end{align}
with 
\begin{align}
        c_{\delta \phi}^{12}(\omega, \bm k) &= \Pi_{11}^A d_{11}^A + \Pi_{12}^A d_{12}^A + \Pi_{21}^A d_{21}^A + \Pi_{22}^A d_{22}^A, \\
        c_{\delta \phi}^{12*}(-\omega^*, -\bm k) &= -\Pi_{22}^A d_{22}^A - \Pi_{12}^A d_{12}^A - \Pi_{21}^A d_{21}^A - \Pi_{22}^A d_{22}^A,
\end{align}
where we have used Eqs. \eqref{RA_Pi} and expressions in \eqref{corrections_RA}. From the relations above we can notice that
\begin{align}
    D_{12}^K(\omega, \bm k) = D_{12}^{K*}(-\omega^*, -\bm k).
\end{align}
where 
\begin{align}
    D_{12}^K (\omega, \bm k)= \frac{\lambda}{4 i} \int_{-\epsilon_c}^{\epsilon_c} d\epsilon d_{12}^K(\epsilon, \omega, \bm k).
\end{align}
Collecting all the necessary terms in Eqs. \eqref{coeffs_Poisson} and their symmetry relations \eqref{n_sym_rel} as well as taking into account Eqs. \eqref{ab_integrated}, \eqref{c_integrated} we can establish the following relations for the charged superconductor
\begin{align}\label{ABC_relations}
    A_{12}^{K*} (-\omega^*, -\bm k) &= A_{12}^{K} (\omega, \bm k), \\
    B_{12}^{K*} (-\omega^*, -\bm k) &= B_{12}^{K} (\omega, \bm k)\\
    C_{12}^{K*} (-\omega^*, -\bm k) &= C_{12}^{K} (\omega, \bm k).
\end{align}
This results in the absence of the amplitude response in the linear response 
\begin{equation}
    F_{\omega_r,\bm k}^a =0.
\end{equation}

\end{widetext}

\bibliography{Magnon_Higgs}

\end{document}